\documentclass[12pt]{article}

\usepackage{placeins}

\usepackage{pdflscape}
\usepackage{amssymb}
\usepackage{amsmath}
\usepackage{amsthm}
\usepackage{cancel}
\usepackage{fullpage}
\usepackage{color}
\usepackage{braket}
\usepackage{graphicx}
\usepackage[small]{subfigure}
\usepackage{multirow}
\usepackage{verbatim}
\usepackage{cite}
\usepackage{bigints}
\usepackage{simplewick}
\usepackage{authblk}
\usepackage{hyperref}
\usepackage{empheq}
\usepackage{tikz}
\usetikzlibrary{calc}
\usepackage{bbm}

\hypersetup{
    linktocpage,
     colorlinks,
     citecolor=darkgreen,
     linkcolor= darkgreen,
     urlcolor=darkgreen
}

\definecolor{darkred}{rgb}{0.5,0.0,0.0}
\definecolor{darkblue}{rgb}{0.0,0.0,0.9}
\definecolor{darkerblue}{rgb}{0.0,0.0,0.5}
\definecolor{purple}{rgb}{0.5,0.0,0.5}
\definecolor{darkgreen}{rgb}{0.0,0.5,0.0}
\definecolor{black}{rgb}{0.0,0.0,0.0}
\definecolor{brown}{rgb}{0.6,0.4,0.2}
\definecolor{newpurple}{rgb}{0.65, 0.38, 0.61}
\definecolor{newyellow}{rgb}{1, 0.75, 0.0}
\definecolor{newblue}{rgb}{0.4, 0.52, 0.85}
\definecolor{newred}{rgb}{0.92, 0.39, 0.21}
\definecolor{newgreen}{rgb}{0.56, 0.69, 0.19}
\definecolor{neworange}{rgb}{0.88, 0.61, 0.14}

\def\be{\begin{equation}}
\def\ee{\end{equation}}

\def\rd{\mathrm{d}}

\def\ecm{E_{\mathrm{CM}}}

\def\muf{\mu}

\def\vb{\bar{v}}

\def\rd{\mathrm{d}}

\def\Mxt{m_X^2}

\newcommand{\jetphox}{{\sc{JetPhox} }}
\newcommand{\peter}{{\sc{PeTeR} }}
\newcommand{\jetphoxd}{{\sc{JetPhox}}}
\newcommand{\peterd}{{\sc{PeTeR}}}

\def\x{x}

\newcommand{\ETg}{\ensuremath{E_{\mathrm T}^{\gamma}}}

\newcommand{\GeV}{\text{GeV}}

\newcommand{\sighatab}{\frac{\rd^2 \hat{\sigma}_{a b}}{\rd w \rd v}}
\newcommand{\sighat}{\frac{\rd^2 \hat{\sigma}}{\rd w \rd v}}

\date{}

\title{Precision direct photon spectra at high energy and comparison to the 8 TeV ATLAS data}

 \author{Matthew D. Schwartz\thanks{schwartz@physics.harvard.edu} }

\affil{\emph{Department of Physics,
Harvard University, Cambridge, MA 02138, USA}}

\begin{document} 
\maketitle

\begin{abstract}
The direct photon spectrum is computed to the highest currently available precision and compared to ATLAS data from 8 TeV collisions at the LHC. The prediction includes threshold resummation at next-to-next-to-next-to-leading logarithmic
order through the program PeTeR, matched to next-to-leading fixed order with fragmentation effects using JetPhox and includes the resummation of leading-logarithmic electroweak Sudakov effects. Remarkably, improved agreement with data can be seen when each component of the calculation is added successively. This comparison  demonstrates the importance of both threshold logs and electroweak Sudakov effects. Numerical values for the predictions are included.
\end{abstract}

\section{Introduction}
The production of energetic photons in the collision of two hadrons is one of the foundational processes of the Standard Model. Photons that come from the collision of two primary partons in the protons are called {\it prompt} or {\it direct}. Due to factorization, the cross section for prompt photons can be computed by convolving non-perturbative, but universal, parton distribution functions (PDFs) with a perturbative partonic cross section. Moreover, by measuring only the photon momentum, inclusive over all other particles, the observable is insensitive
to hadronization effects and therefore particularly clean. Thus direct photon production has provided one of the best tests of the Standard Model at hadron colliders over the last thirty years. In fact, the precision by which its spectrum can be predicted allows for unique sensitivity into physics beyond the Standard Model.
This paper reports on the state-of-the art theory calculation and comparison to recent data from the Large Hadron Collider.

The theoretical calculation of the photon spectrum at leading order (order $\alpha_s$) is straightforward. At next-to-leading
order (NLO) in Quantum Chromodynamics (QCD), the result has been known since the early 1980s~\cite{Aurenche:1983ws,Aurenche:1987fs,Gordon:1993qc}.
The NLO photon spectrum, inclusive over all hadrons, is available in  analytic form at the parton level, however it must be integrated numerically
against the PDFs to produce the observable cross section. 
A number of computer codes are available to produce this NLO cross section, including 
\jetphox  \cite{Catani:2002ny}  and \peter~\cite{Becher:2013vva,peter}.
The full result at next-to-next-to-leading order (NNLO) is not yet known, although the technology to complete it is available.
For example, the NNLO distributions for the analagous processes $W$ and $Z$ boson production were completed recently~\cite{Boughezal:2016dtm,Boughezal:2015ded}. 

An alternative to computing the spectrum order-by-order in $\alpha_s$ is to compute some of the terms to a given order and other terms to all order in $\alpha_s$. 
Naturally, the efficacy of such an approximation revolves around which terms are included and why they should be more
valuable than the terms that are neglected. For the direct photon spectrum, the relevant physical scales are the machine center-of-mass energy $\sqrt{S} = \ecm = 8$ TeV and the energy (or transverse energy $E_T=|\vec{p}_T|$) of  the photon. In the {\it threshold limit}, when $E_T \to \ecm/2$, the kinematics only allows for the photon to be recoiling against a single collimated jet. Indeed, if we denote everything other than the photon in the event as $X$, then, by pure kinematical considerations, the mass of $X$, $M_X = \sqrt{p_X^2}$, must go to zero as $E_T \to \ecm/2$ and the energy of $X$ must also approach $\ecm/2$. Thus $X$ must look like a jet. As the jet mass translates directly into $E_T$, we can then use the domination of the mass by soft and collinear physics, which are well understood in QCD, to see that the photon $E_T$ spectrum is also dominated by soft and collinear physics. Including the associated large  logarithms, resummed to all orders, leads to a  a precision calculation beyond the NLO 
level.

For direct photon production, the resummation of large logarithms was done to next-to-leading logarithmic order (NLL)~\cite{Laenen:1998qw,Catani:1998tm,Catani:1999hs} in the late 1990s. Partial higher order results were soon-after produced using soft-gluon resummation~\cite{Kidonakis:1999hq,Kidonakis:2003bh}.
The complete resummation at NNLL and N${}^3$LL was achieved using Soft-Collinear Effective Theory (SCET) \cite{Bauer:2000yr,Bauer:2001yt,Beneke:2002ph} overestimatesthe past few years. The relevant
factorization theorem was derived in~\cite{Becher:2009th} and applied to photon and $W$ and $Z$ production in~\cite{Becher:2011fc,Becher:2012xr}. 
Additional ingredients were computed in~\cite{Becher:2012za,Becher:2010pd,Becher:2013vva}.
These papers achieved the resummation at the next-to-next-to-next-to-leading logarithmic level (NNNLL). The calculation has been implemented in the public computer
code \peter~\cite{peter}. 

One complication of the photon $E_T$ spectrum, compared to say, the $Z$ boson spectrum, is that one cannot easily tell experimentally whether the observed photons were prompt or not. A significant background comes from the decay of $\pi_0$ particles. This fragmentation contribution can be modeled and tuned to data. Nevertheless, it can overwhelm the signal, diminishing the appealing features of direct photon production. The standard approach to dealing with $\pi_0$ decays is to require the observed photon to be isolated. 
The idea is that if there are $\pi_0$'s decaying to photons, there will most likely be other hadrons nearby the $\pi_0$, while prompt photons are naturally isolated. 
In~\cite{Aad:2016xcr} the isolation requirement is that
\be
E_T^{\text{iso}} < 4.8~ \GeV + 0.0042 ~E_T
\ee
where $E_T^{\text{iso}}$ is the total energy not in the photon in a cone of radius $R=0.4$ around the photon. 

To incorporate the isolation requirement into the theory prediction one must also account for the fact that in addition to reducing the background
the isolation requirement also affect the direct photon signal. The (positive) contribution to the cross section from the fragmenting hadrons passing the isolation criteria and the (negative) contribution to the cross section from direct photons failing the isolation are included in the program \jetphox. An important observation is that at asymptotically high $p_T$, both effects become negligible: the fragmentation correction is a power corrections in $\frac{\Lambda_{\text{QCD}}}{E_T}$. This, the connection to beyond-the-Standard-Model physics and the relative importance of the resummation, motivates focusing on very high $E_T$ photons. 

A final theoretical ingredient for a precision prediction are electroweak corrections, for example from loops of photons or $W$ bosons connecting the charged quarks involved in the partonic process. Such loops can generate large logarithms near threshold, called {\it electroweak Sudakov logs}. The analysis of electroweak Sudakov logs in the context of direct photon was recently done in~\cite{Becher:2013zua,Becher:2015yea}. The effect of including these logs is to lower the direct photon cross section at high $E_T$ by up to around 10\%. 

This paper provides a numerical prediction to the highest currently available precision of the isolated direct photon spectrum. The predictions are binned in rapidity according to the recent ATLAS measurement~\cite{Aad:2016xcr}. The results of the calculations described in
this paper were given to ATLAS in a private communication and have already been used in the experimental publication. This paper describes how the calculation was performed and tabulates the results. Theory predictions are also included at intermediate levels of precision, so that the importance of electroweak and threshold logarithms can be separately seen.

\section{Calculational details}
As discussed above, the theoretical prediction of the direct photon spectrum at NLO has been known for some decades. The fragmentation and isolation criteria are included using the program \jetphox. The \jetphox results in this paper, including the central values,
scale uncertainty and PDF uncertainty were produced by ATLAS.  We have not attempted to reproduce them. Instead, we supplement the \jetphox
results with the N${}^3$LL threshold resummation using \peter and the electroweak Sudakov effects.

For threshold resummation, the starting point is the factorization formula~\cite{Gordon:1993qc}
\begin{equation}
\frac{\rd^2 \sigma}{\rd y \rd p_T} 
=  \frac{2}{p_T}  \sum_{ab}
\int^{1 -  \frac{p_T}{\ecm} e^{- y}}_{\frac{p_T}{\ecm} e^y} \rd v 
\int_{\frac{p_T}{\ecm} \frac{1}{v} e^y}^1 \rd w 
\left[ \x_1 f_{a/N_1} (\x_1, \mu) \right] \left[ \x_2 f_{b/N_2} (\x_2, \mu) \right]
\sighatab \, , \label{csec}
\end{equation}
where the sum is over the different partonic channels. Here $w$ and $v$ are partonic variables defined in terms of the usual $2\to2$ Mandelstam variables as
\begin{equation}
v = 1 + \frac{\hat{t}}{\hat{s}}
\, , 
\hspace{1em}
w = -  \frac{\hat{u}}{\hat{s} + \hat{t}}\,.
\end{equation}
Using $v$ and $w$ rather than $\hat{s}$ and $\hat{t}$ improves the convergence of the integrals but is not strictly necessary. 

For the direct photon cross section, one can proceed to compute $\sighatab$ order by order in $\alpha_s$. At leading order, 
\begin{equation}
\sighatab 
=\frac{\vb}{p_T^2} \, \widetilde\sigma_{ab}(v) \,\delta(1-w)
\end{equation}
where the fiducial cross section is slightly different in the annihilation ($q \bar q \to g \gamma$)
\be
 \widetilde\sigma_{q \bar q}(v) = \pi \alpha_{\mathrm em} e_q^2 \alpha_s (\mu) \frac{2 C_F}{N_c}  \left( v^2 + \vb^2 \right)\frac{1}{\vb}\,,
\ee
and Compton ($q g \to q \gamma$)
\be
 \widetilde\sigma_{q g} (v) = \pi \alpha_{\mathrm em} e_q^2 \alpha_s (\mu) \frac{1}{N_c}\nonumber
\ee
channels. 

While at leading order $\rd \hat\sigma \sim \delta(1-w)$, at higher orders $\rd \sigma$ has logarithms $\ln (1-w)$. These are the large logarithms which can be resummed. To perform the resummation, the threshold expansion uses the following factorization formula:
\begin{equation}\label{sigmafact}
\sighat
=w\, \widetilde\sigma(v)\, H (p_T, v ,\mu) \int \rd k\, J (\Mxt - (2 E_J) k,\mu)\, S (k,\mu) \, .
\end{equation}
with $H$ the hard function, $J$ the jet function, and $S$ the soft function. Operator definitions of these functions can be found in~\cite{Becher:2009th}. 
The large logarithms are resummed by evaluating these functions to order $\alpha_s^2$ at each of their respective scales $\mu_h$, $\mu_j$ and $\mu_s$, then evolving the functions to a common scale $\mu$ using renormalization group evolution. The common scale is taken to be $\mu=\mu_f$, the factorization scale where the PDFs are evaluated. 

A satisfying observation about the resummed expression is that the hard, jet, and soft functions can reveal their own natural scales~\cite{Becher:2007ty}. The observation is that, in a NLO calculation the dependence on the renormalization group scale $\mu$ is typically monotonic. It is natural to choose $\mu=E_T$ for direct photon, but this choice is essentially arbitrary. On the other hand, if we include, the hard function only in Eq.~\eqref{sigmafact}, one finds that the cross section has a maximum at some value $\mu=\mu_h$. Taking $\mu=\mu_h$ then is a natural choice for minimal scale sensitivity. Similarly, the cross section including only the jet function has a minimum at $\mu=\mu_j$,
and the soft function a maximum at $\mu=\mu_s$. The contrast between NLO and the effective field theory calculations arises because the NLO calculation sets the hard, jet and soft scales equal. By separating the different modes, the arbitrariness of the scale choice at fixed order is removed. In~\cite{Becher:2012xr} numerical fits were performed to the location of the maxima and minima for the soft, jet and hard functions. The result are the default scales in \peter.

For the direct photon calculation we take $\mu_h = \mu_f = E_T$, as this is the default scale in \jetphox and used by ATLAS. For the jet scale we take the fit result~\cite{Becher:2012xr}
\be
\mu_j = \frac{7}{12}E_T\left(1-2\frac{E_T}{\ecm}\right)   
\label{mujchoice}
\ee
and we take the natural seesaw scale $\mu_s = \frac{\mu_j^2}{\mu_h}$ for the soft function~\cite{Schwartz:2007ib}. That is, in addition to Eq.~\eqref{mujchoice} we take
\begin{align}
\mu_h &= E_T\\
\mu_f &= E_T\\
\mu_s &= \frac{\mu_j^2}{\mu_h}
\end{align}

To produce a result which is accurate to NLO, includes the fragmentation and isolation effect, and also includes the higher order terms computed from threshold resummation, we compute~\cite{Becher:2007ty}
\begin{equation}\label{match}
\left(\frac{\rd^2\sigma}{\rd v \rd w}\right)^{\text{\peter+\jetphox}} 
= 
\left(\frac{\rd^2\sigma}{\rd v \rd w}\right)^{\text{\peter}}   
- \left(\frac{\rd^2\sigma}{\rd v \rd w}\right)^{\text{\peter}}_{\mu_h=\mu_j=\mu_s=\muf}
+\left(\frac{\rd^2\sigma}{\rd v \rd w}\right)^{\text{\jetphox}}
\,.
\end{equation}
The second term on the right-hand side subtracts off the fixed order expansion of \peterd. By setting all the scales equal, this term has all resummation turned off. The fixed order contribution is added back in by the final term, including also the fragmentation and isolation contribution.

\begin{figure}[th!]
\begin{center}
\begin{tikzpicture}[scale=1.0]
 	\coordinate (c1) at (-1,0);
 	\coordinate (c2) at (8,0);
 	\coordinate (c3) at (-1,-6);
 	\coordinate (c4) at (8,-6);
 	\coordinate (c5) at (-1,-12);
 	\coordinate (c6) at (8,-12);
 	\coordinate (dc1) at (0.5,1.8);
 	\coordinate (dc2) at (0.5,2.5);
 	\coordinate (dc3) at (0.5,2.4);
        \coordinate (dc4) at (-1.4,-1.0);
        \coordinate (dc5) at (-1.4,-1.4);
        \coordinate (dc6) at (2.5,-1.5);
        \node at (c1) {\includegraphics[width=0.48\columnwidth]{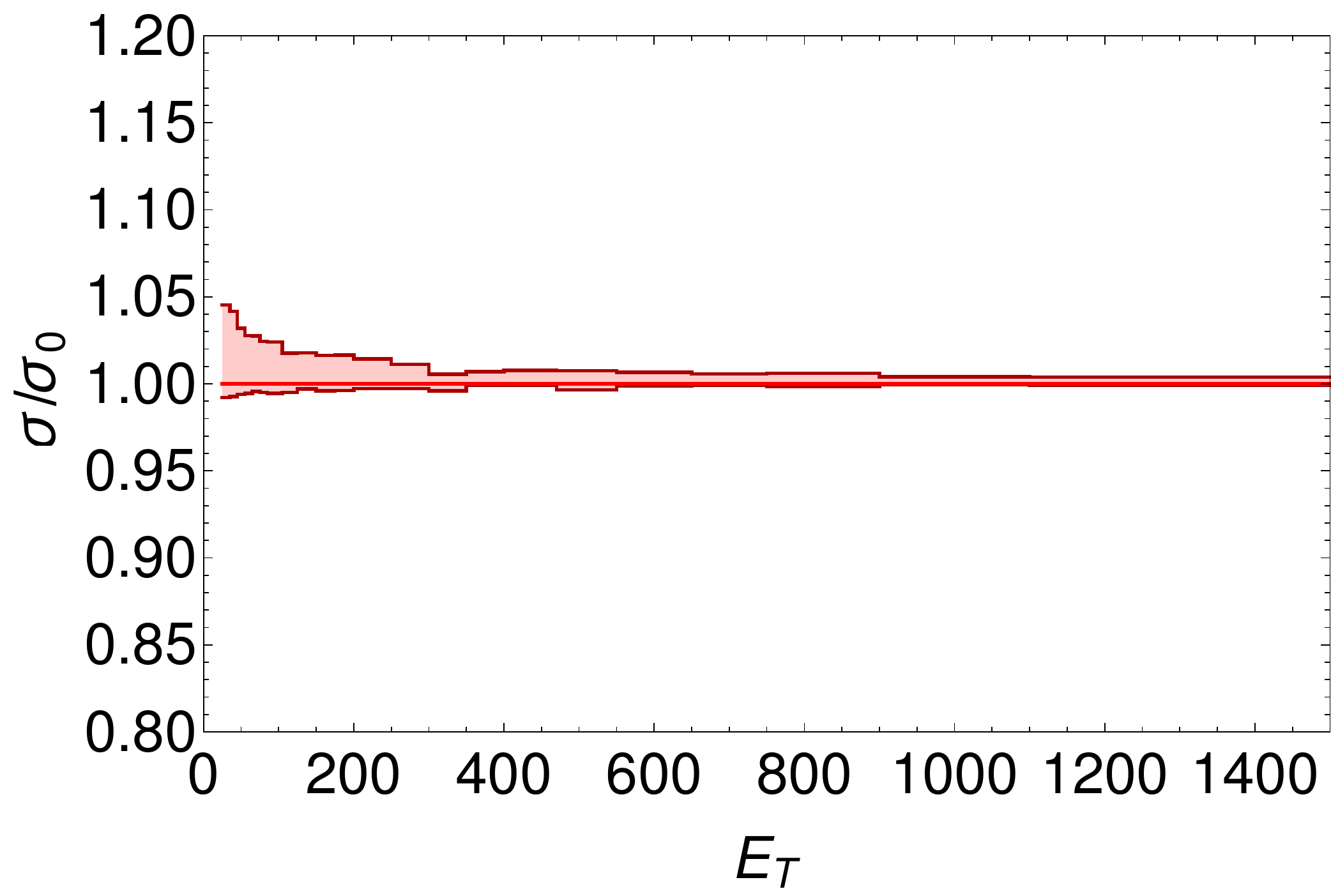}};
        \node at (c2) {\includegraphics[width=0.48\columnwidth]{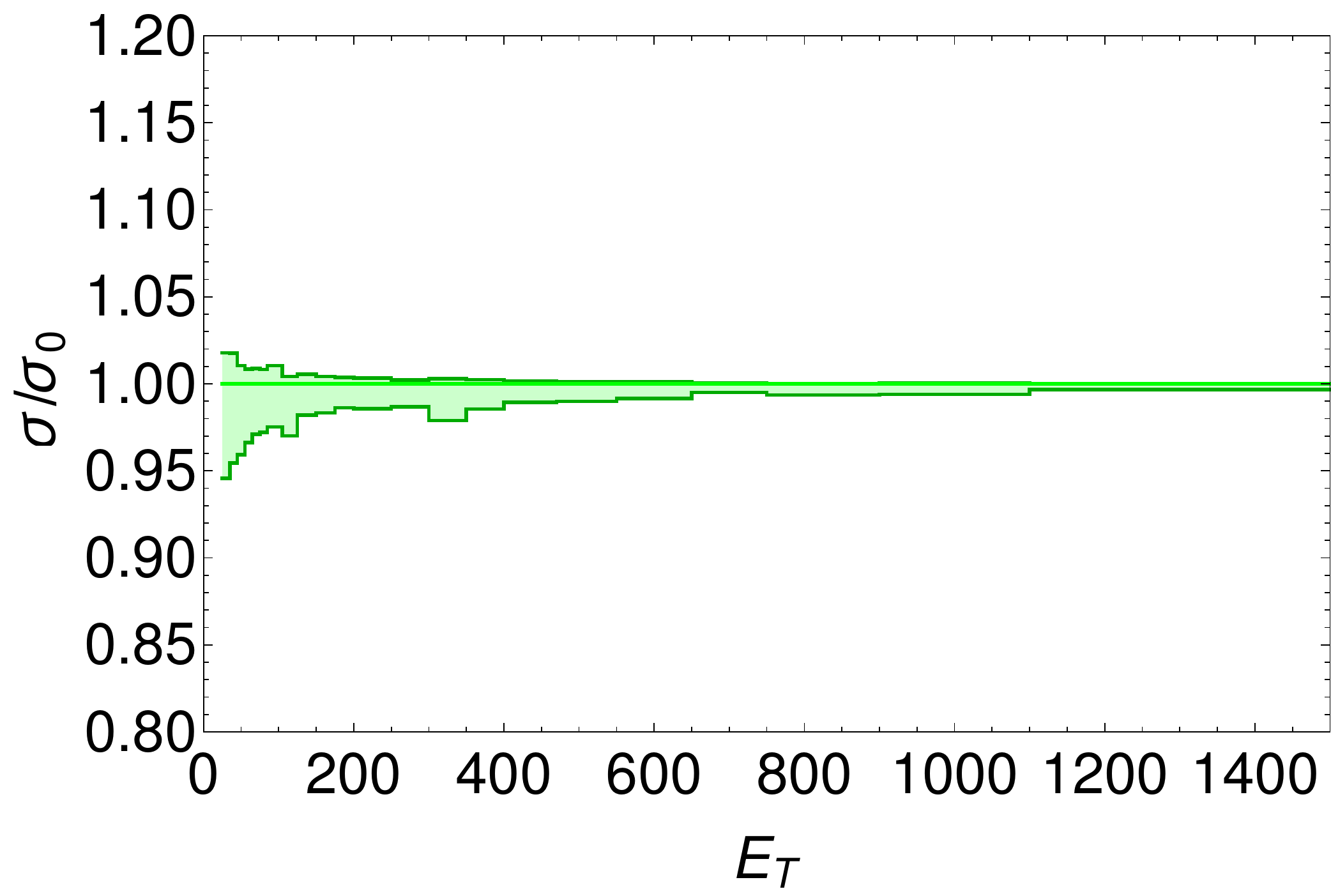}};
        \node at (c3) {\includegraphics[width=0.48\columnwidth]{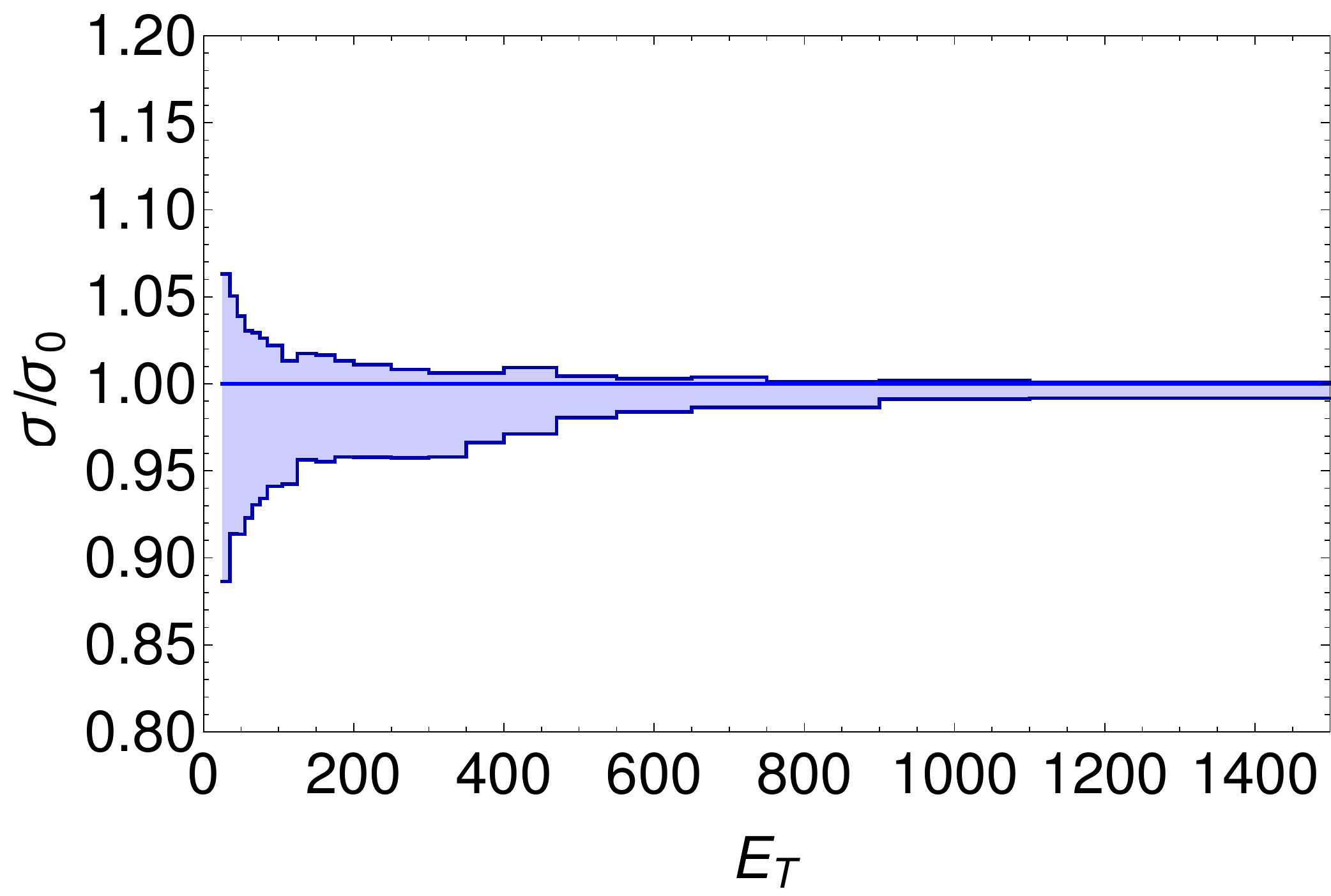}};
        \node at (c4) {\includegraphics[width=0.48\columnwidth]{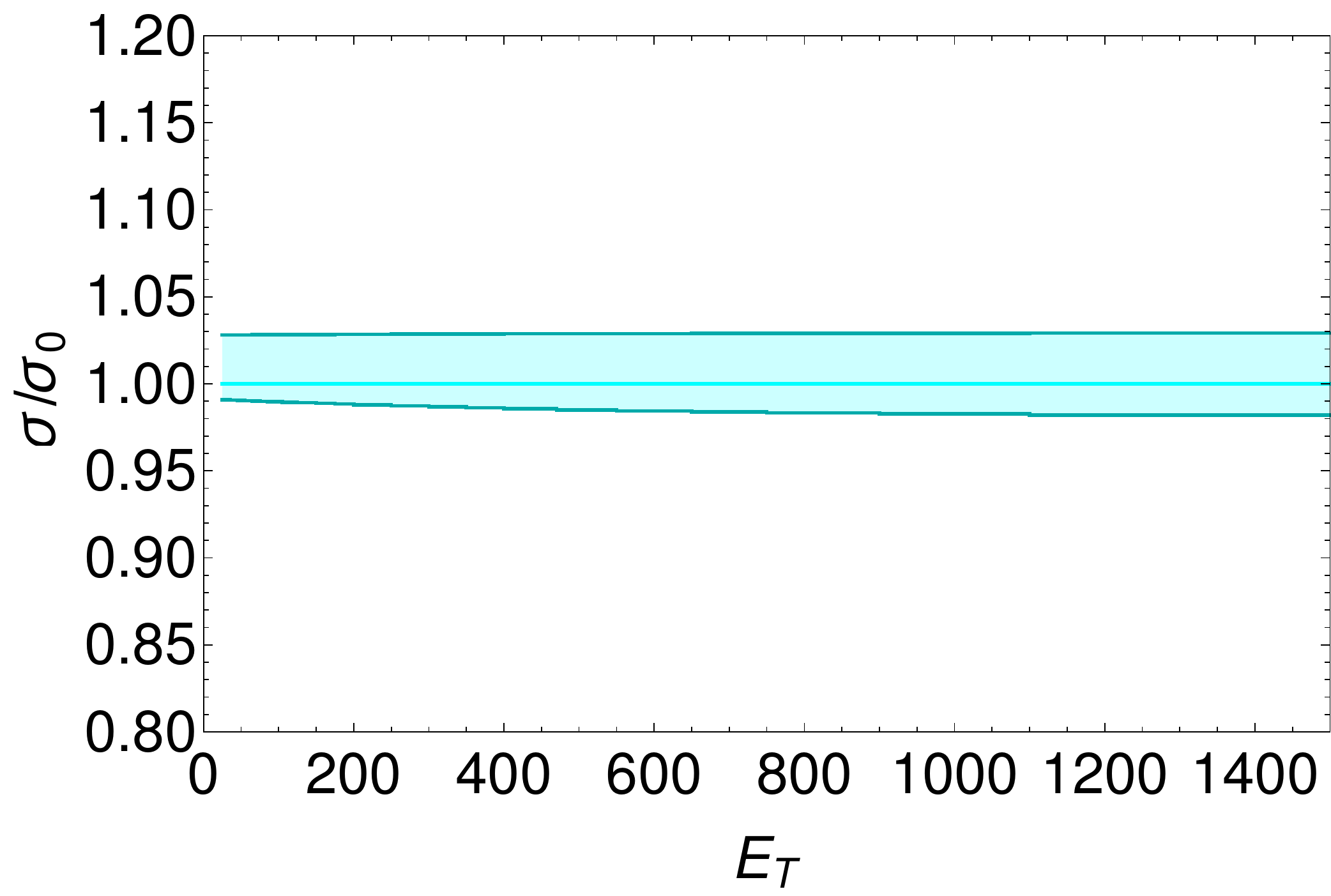}};
        \node at (c5) {\includegraphics[width=0.48\columnwidth]{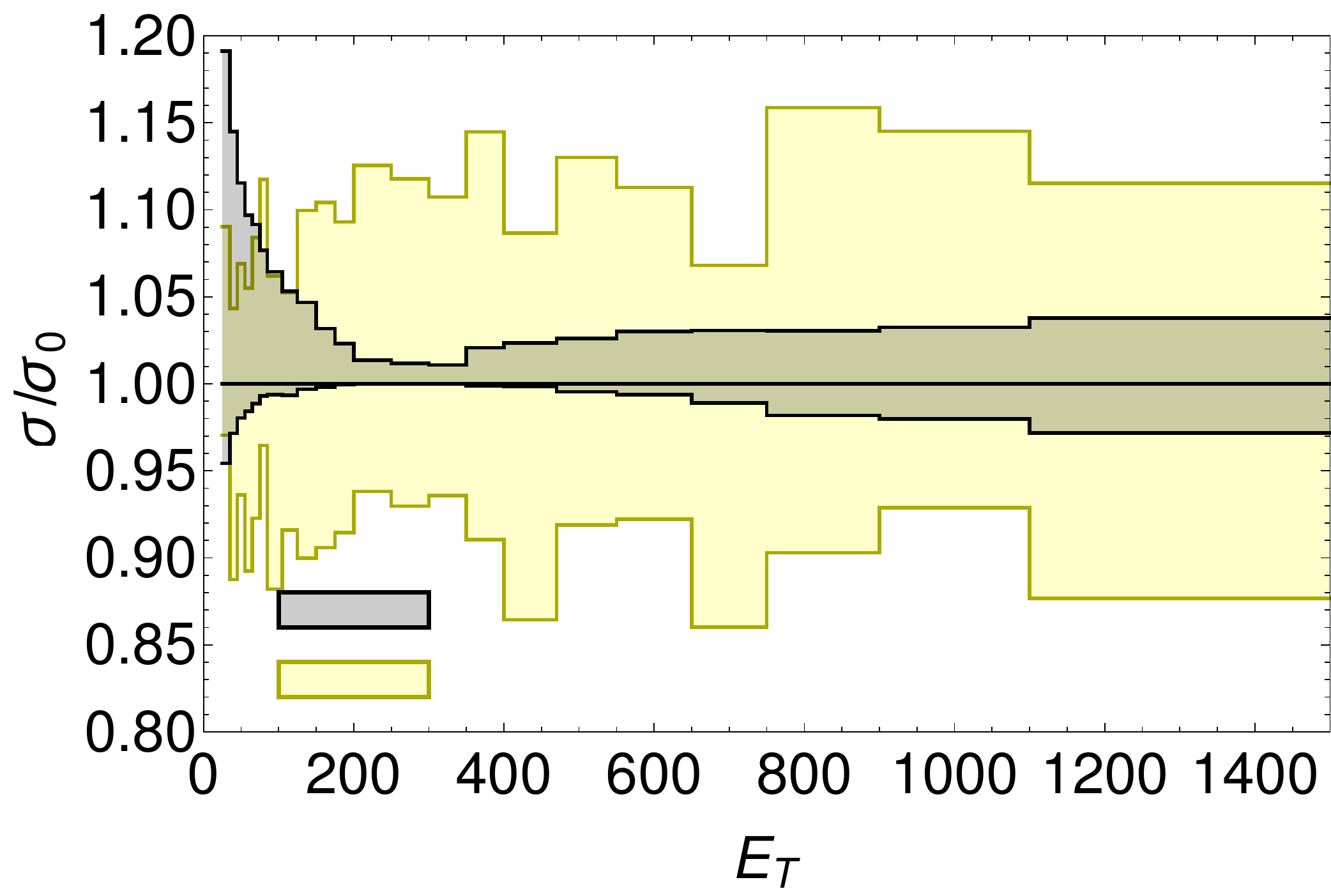}};
        \node at (c6) {\includegraphics[width=0.48\columnwidth]{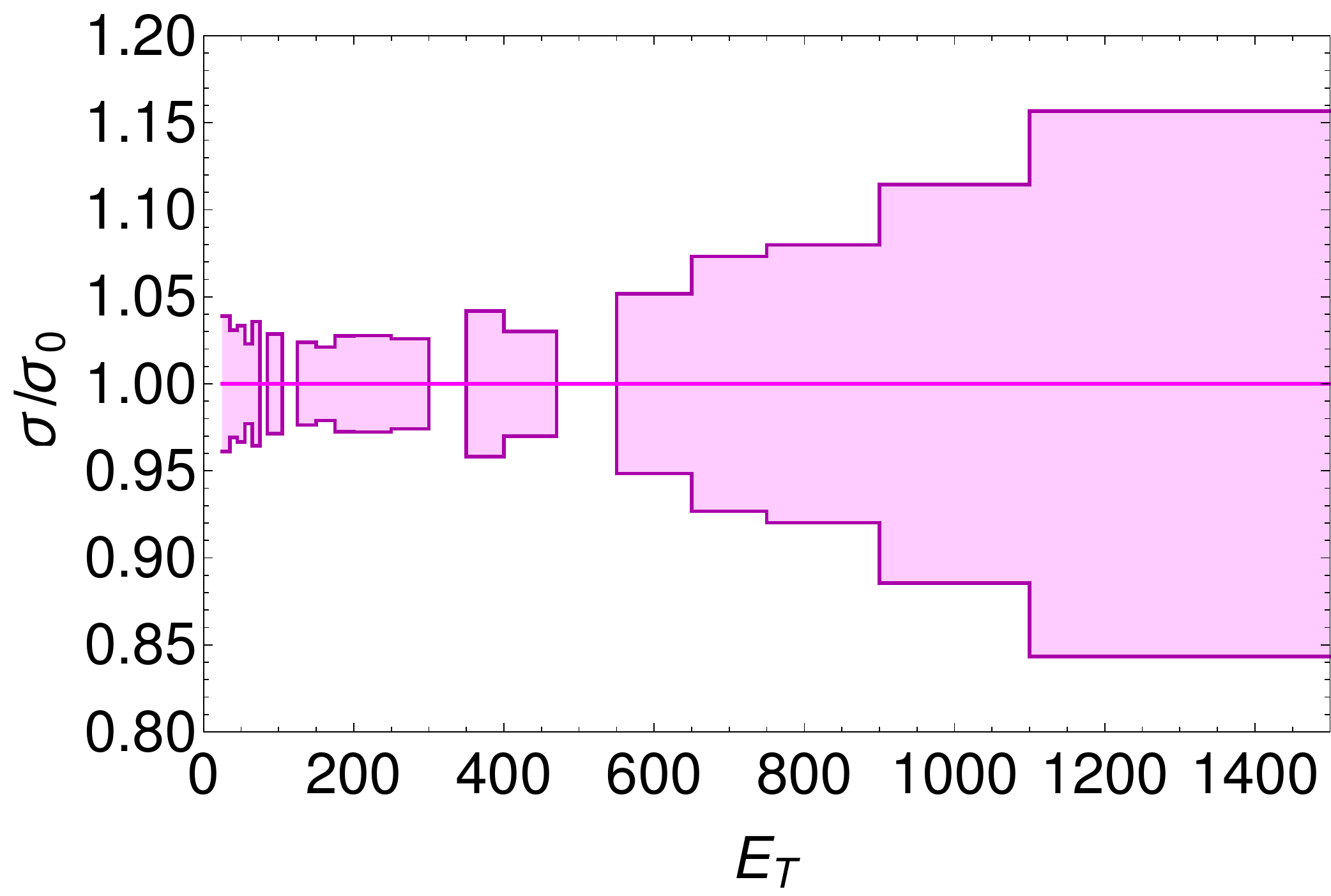}};
	\node[below,scale=1.0] at ($(c1)+(dc1)$) {Hard scale uncertainty};
%	\node[above,scale=1.0] at ($(c1)+(dc2)$) {$|\eta| < 0.6$};

	\node[below,scale=1.0] at ($(c2)+(dc1)$) {Jet scale uncertainty};
%	\node[above,scale=1.0] at ($(c2)+(dc2)$) {$|\eta| < 0.6$};

	\node[below,scale=1.0] at ($(c3)+(dc1)$) {Soft scale uncertainty};
%	\node[above,scale=1.0] at ($(c3)+(dc2)$) {$|\eta| < 0.6$};

	\node[below,scale=1.0] at ($(c4)+(dc1)$) {Electroweak uncertainty};

	\node[below,scale=1.0] at ($(c5)+(dc2)$) {Factorization scale uncertainty};
	\node[right,scale=0.6] at ($(c5)+(dc5)$) {{\bf JetPhox}};
	\node[right,scale=0.6] at ($(c5)+(dc4)$) {{\bf JetPhox + PeTeR}};

	\node[below,scale=1.0] at ($(c6)+(dc3)$) {PDF uncertainty};
%	\node[above,scale=1.0] at ($(c4)+(dc2)$) {$|\eta| < 0.6$};
\end{tikzpicture}
\caption{Theoretical uncertainties. The hard, jet and scale uncertainties come from varying the scales by a factor of 2 around their default values. The electroweak uncertainty is taken from~\cite{Becher:2015yea}. The PDF uncertainty is taken from ATLAS~\cite{Aad:2016xcr}, who computed it using \jetphoxd. Note from the bottom left panel that  by matching to the resummed distribution, the factorization scale uncertainty of \jetphox is severely reduced.}
\label{fig:unc}
\end{center}
\end{figure}
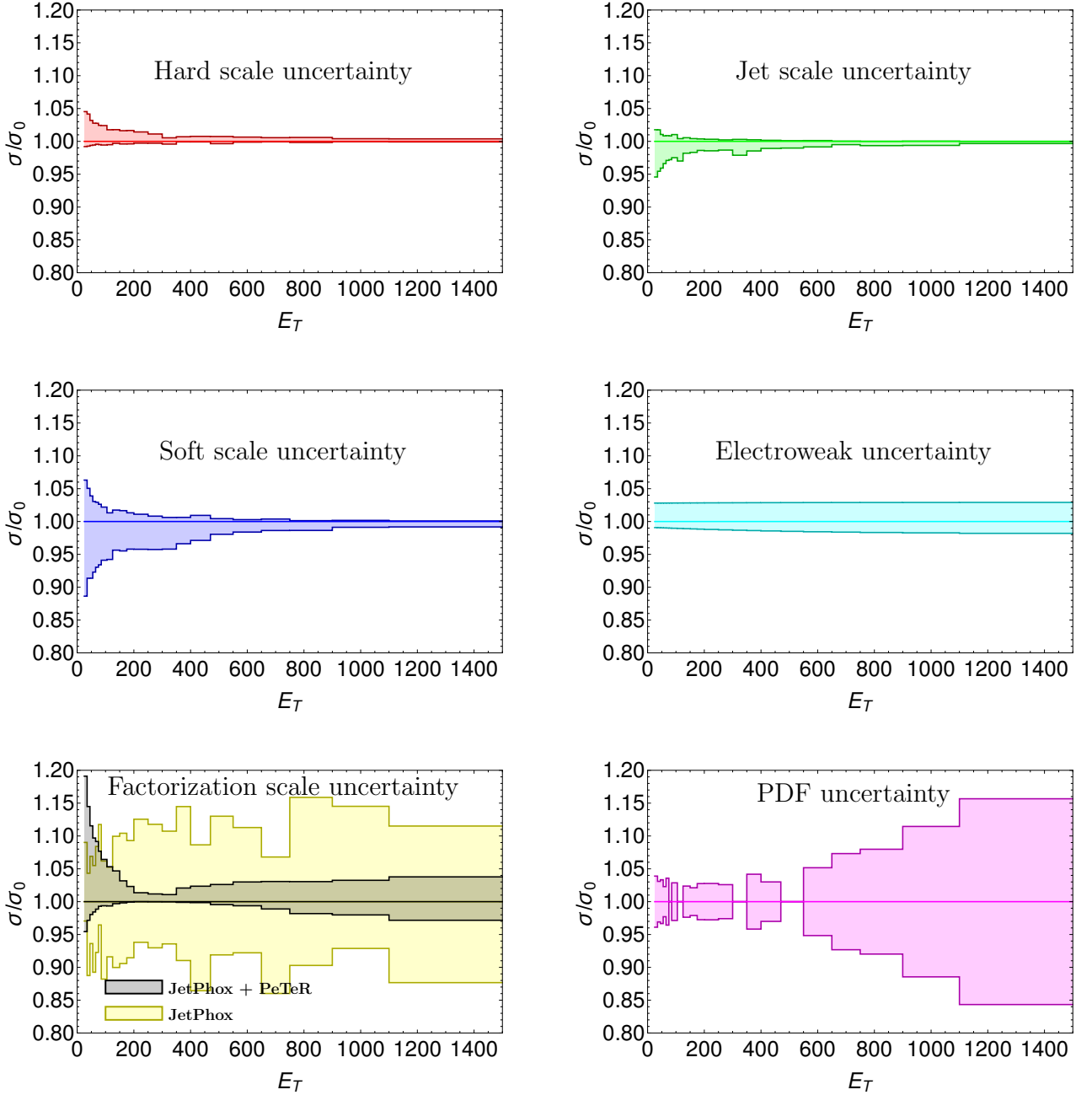

The {\it matching} option in \peter uses this approach but with \jetphox replaced by the purely perturbative NLO result. However, because we want to include fragmentation and to match to \jetphox rather than NLO, we match separately. That is, we run \peter nine times for each $E_T$ and rapidity bin: once for the central value, and two for each scale variation:
%The scale variations are computed by varying 
the hard, jet, soft and factorization scales are independently varied by factors of 2. 
For example, for the hard scales, we take $\mu_h = 2 E_T$, $\mu_h = E_T$ and $\mu_h = \frac{1}{2}E_T$. Because the default scales are chosen to be extrema of the scale variation, we cannot then estimate the uncertainty simply by comparing the $\mu_h = 2 E_T$ and $\mu_h = \frac{1}{2}E_T$ results. Instead, we fit a quadratic function of $\ln \mu$ to the 3 fit values and take the maximum and minimum of this function along the variation region. 
The separate variations are shown in Fig.~\ref{fig:unc}. 

There are a couple of things things worth noting from Fig.~\ref{fig:unc}. First, we see that the factorization scale uncertainty from \jetphox is significantly reduced by this matching procedure. Second, we see that the PDF uncertainty is by far the dominant uncertainty at high $E_T$. This is good, because it means that the precision comparison between theory and data of this observable can be used to improve PDF fits. In particular, high $E_T$ corresponds to $x \sim 1$ where the PDF uncertainties are relatively large.

For the electroweak corrections we take the results from~\cite{Becher:2015yea}. The correction is fit by a smooth function. For $\ecm =  8$ TeV, this function using the central values of the scale choices is
\be
\sigma \to \sigma \frac{1.713 - 21.68 x + 12.16 x^2 - 3.05 x^3}{1 - 0.023355 y + 0.001231 y^2}
\ee
where
\be
x =\frac{E_T}{1~\text{TeV}},\qquad  
%y = \sqrt\frac{1}{7}
y = \frac{ \sqrt{8~\text{TeV}} - \sqrt{7~\text{TeV}} }{ \sqrt{7~\text{TeV}}}
\ee
Fits for the scale variations can be found in~\cite{Becher:2015yea}. The electroweak scale uncertainty is shown in the bottom-right panel of Fig.~\ref{fig:unc}. 

It is worth noting that the entire direct photon cross section is directly proportional to $\alpha_e(\mu)$. Including electroweak effects at
leading order, any choice of $\mu$ is as good as any other -- varying $\mu$ can be compensated for with NLO terms. However, the difference
between $\alpha_e = \frac{1}{137} \approx 0.0073$, corresponding to $\mu \lesssim m_e$, and $\alpha_e(m_Z) \approx \frac{1}{129} \approx 0.0078$ is a 6\% effect, easily observable. Including the electroweak Sudakov effects is therefore critical to lessening this scale sensitivity.
If only leading-order in $\alpha_e$ results are available, one can try to choose $\mu$ to approximate the correct, resummed result. From 
both theoretical arguments\cite{Czarnecki:1998tn,Becher:2013zua,Becher:2015yea} and by comparison to data, it seems clear that taking  $\alpha_e = \frac{1}{137}$, the default
for \jetphox and most NLO numerical calculations is not appropriate for the direct photon spectrum.

  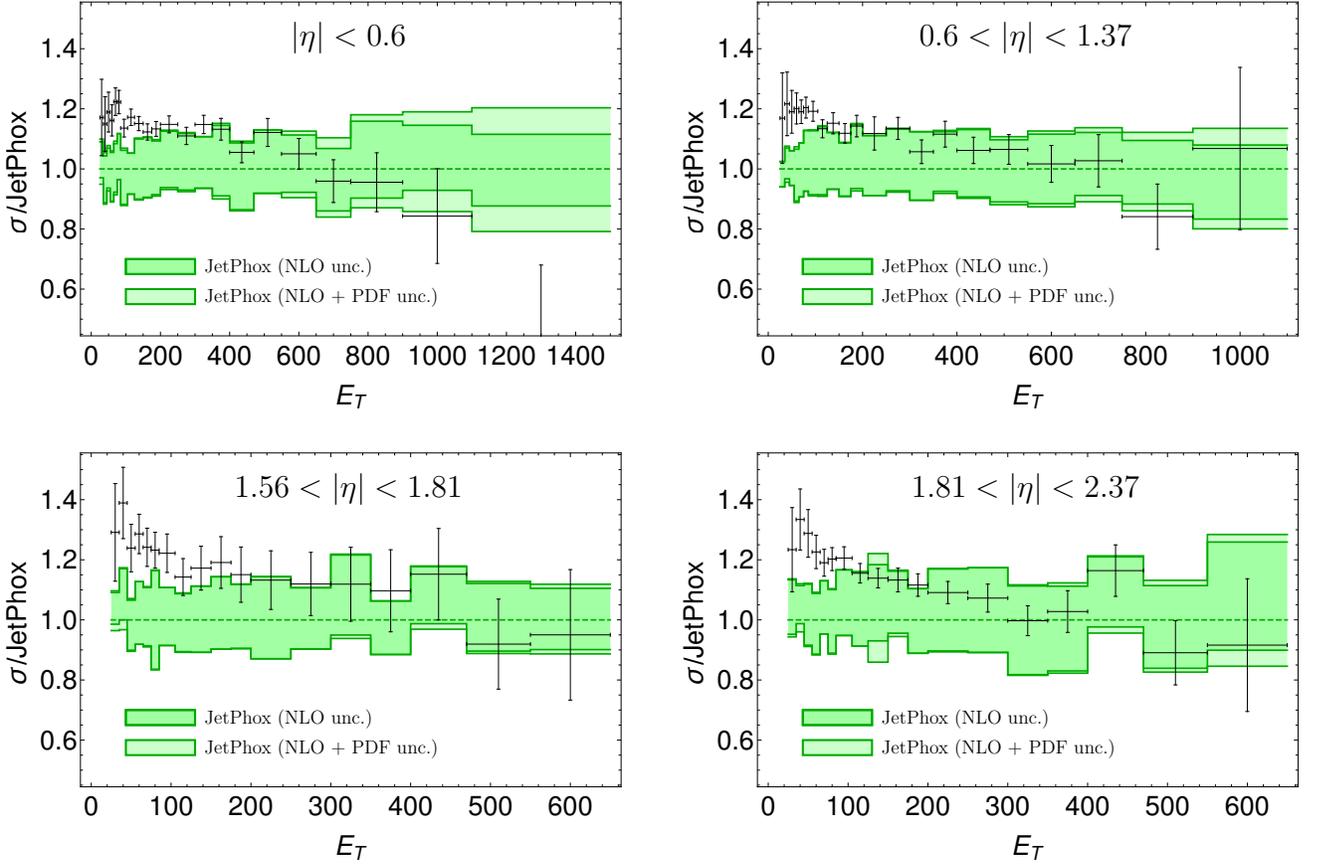
\begin{figure}[t]
\begin{center}
\begin{tikzpicture}
 	\coordinate (c1) at (-1,0);
 	\coordinate (c2) at (8,0);
 	\coordinate (c3) at (-1,-6);
 	\coordinate (c4) at (8,-6);
%  	\coordinate (dc1) at (-1.5,1.8);
%         \coordinate (dc2) at (0.8,2.2);
%         \coordinate (dc3) at (0.8,1.8);
 	\coordinate (dc1) at (0.5,1.8);
        \coordinate (dc2) at (-1.5,-0.9);
        \coordinate (dc3) at (-1.5,-1.3);
        \node at (c1) {\includegraphics[width=0.5\columnwidth]{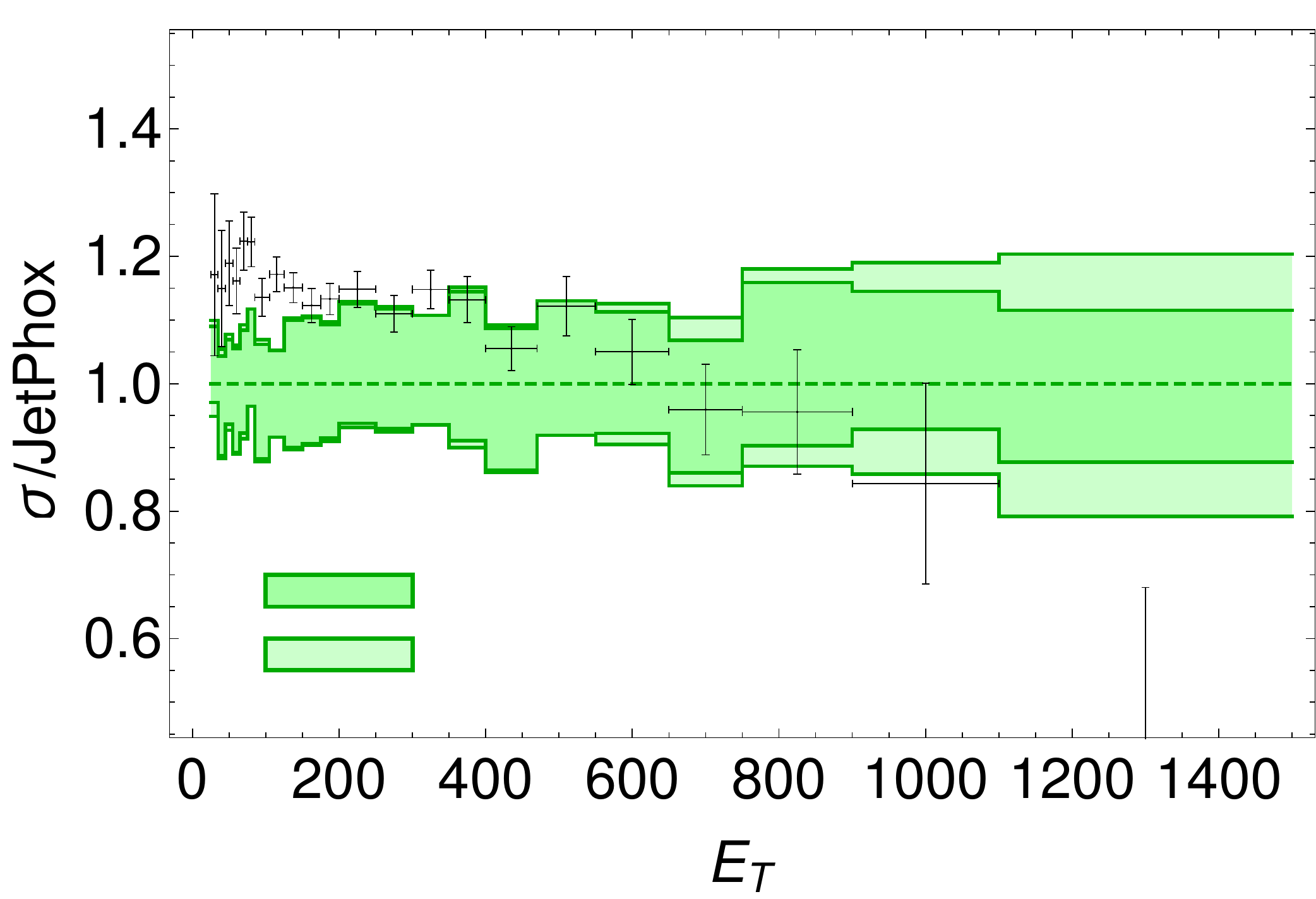}};
        \node at (c2) {\includegraphics[width=0.5\columnwidth]{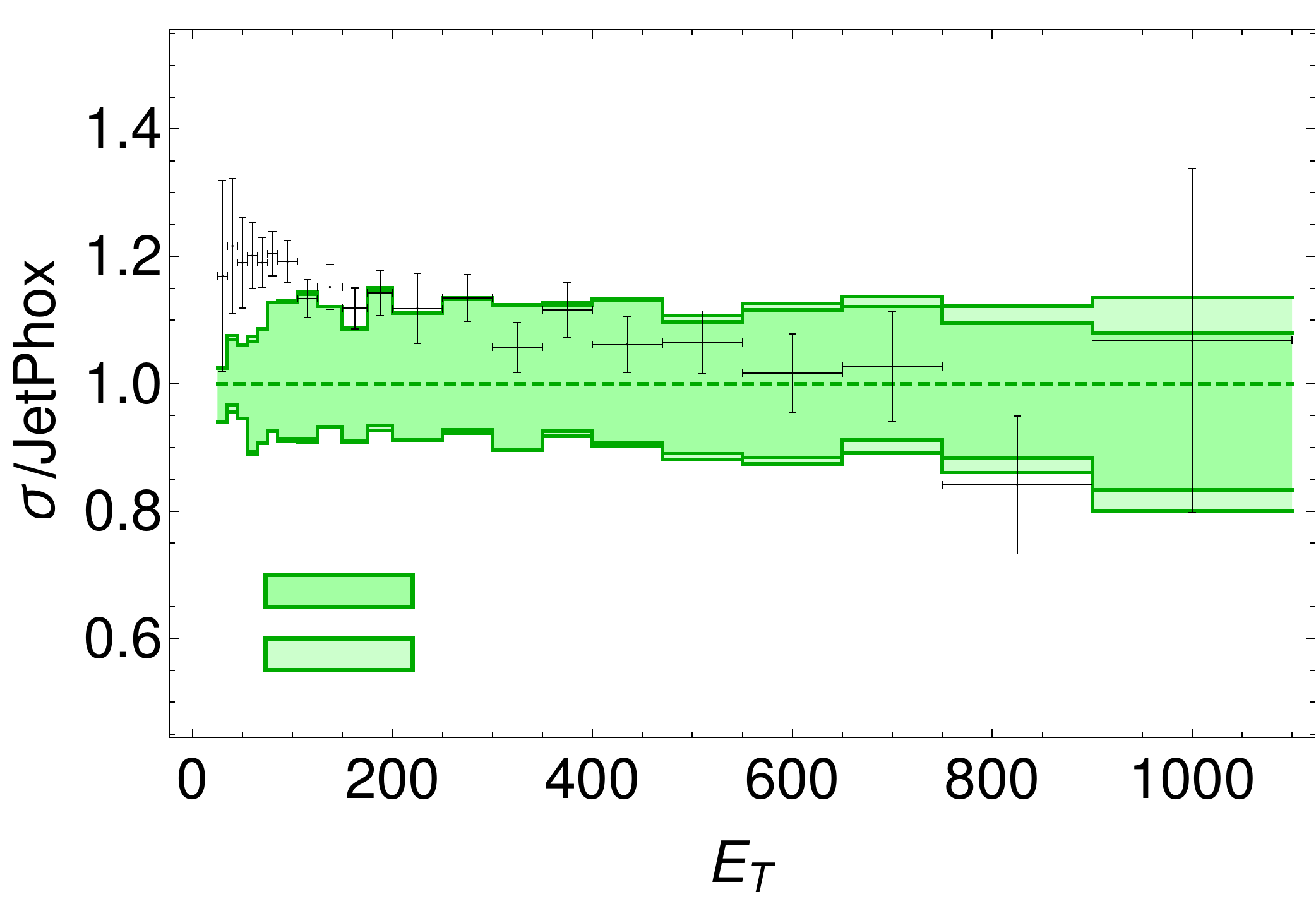}};
        \node at (c3) {\includegraphics[width=0.5\columnwidth]{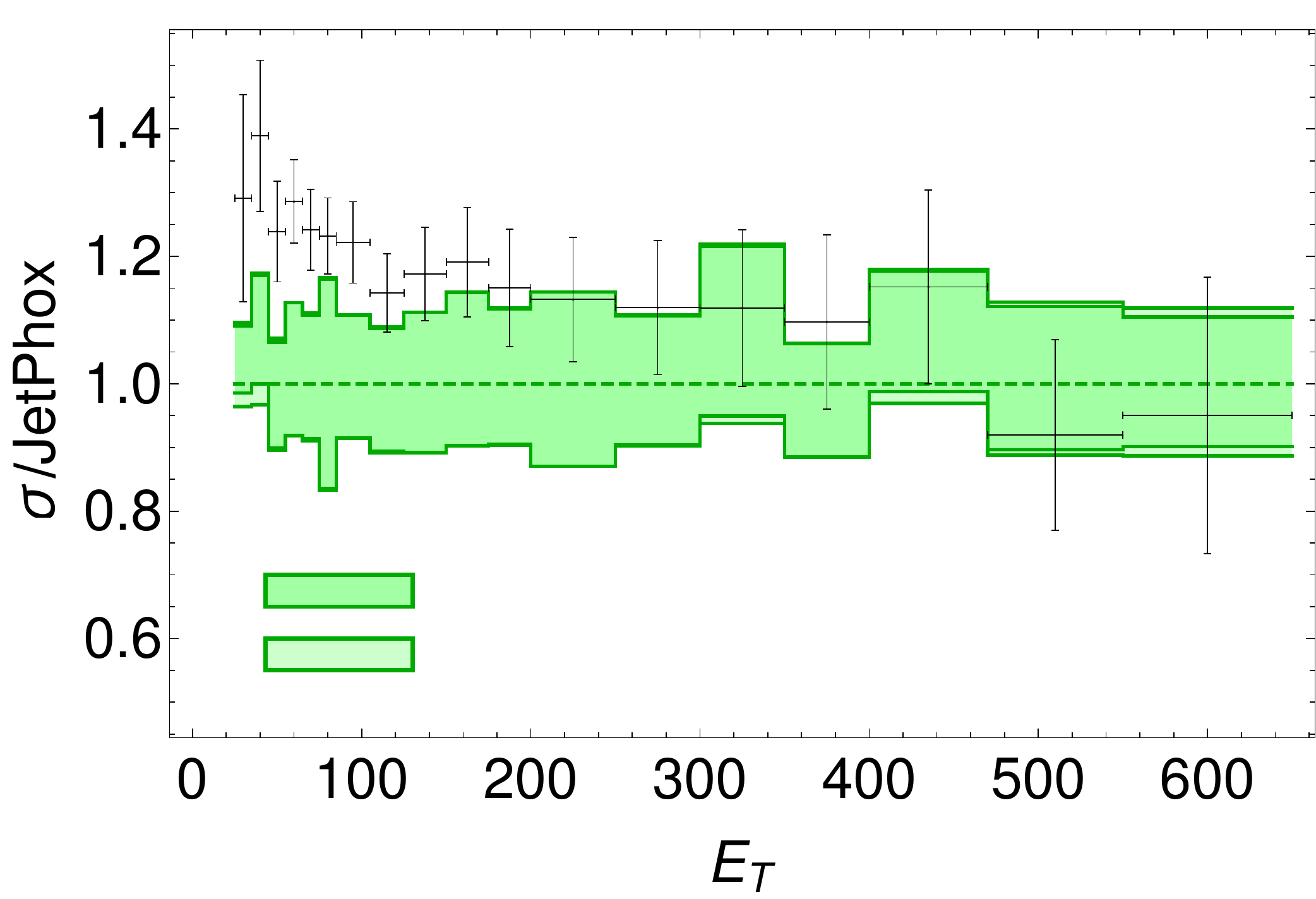}};
        \node at (c4) {\includegraphics[width=0.5\columnwidth]{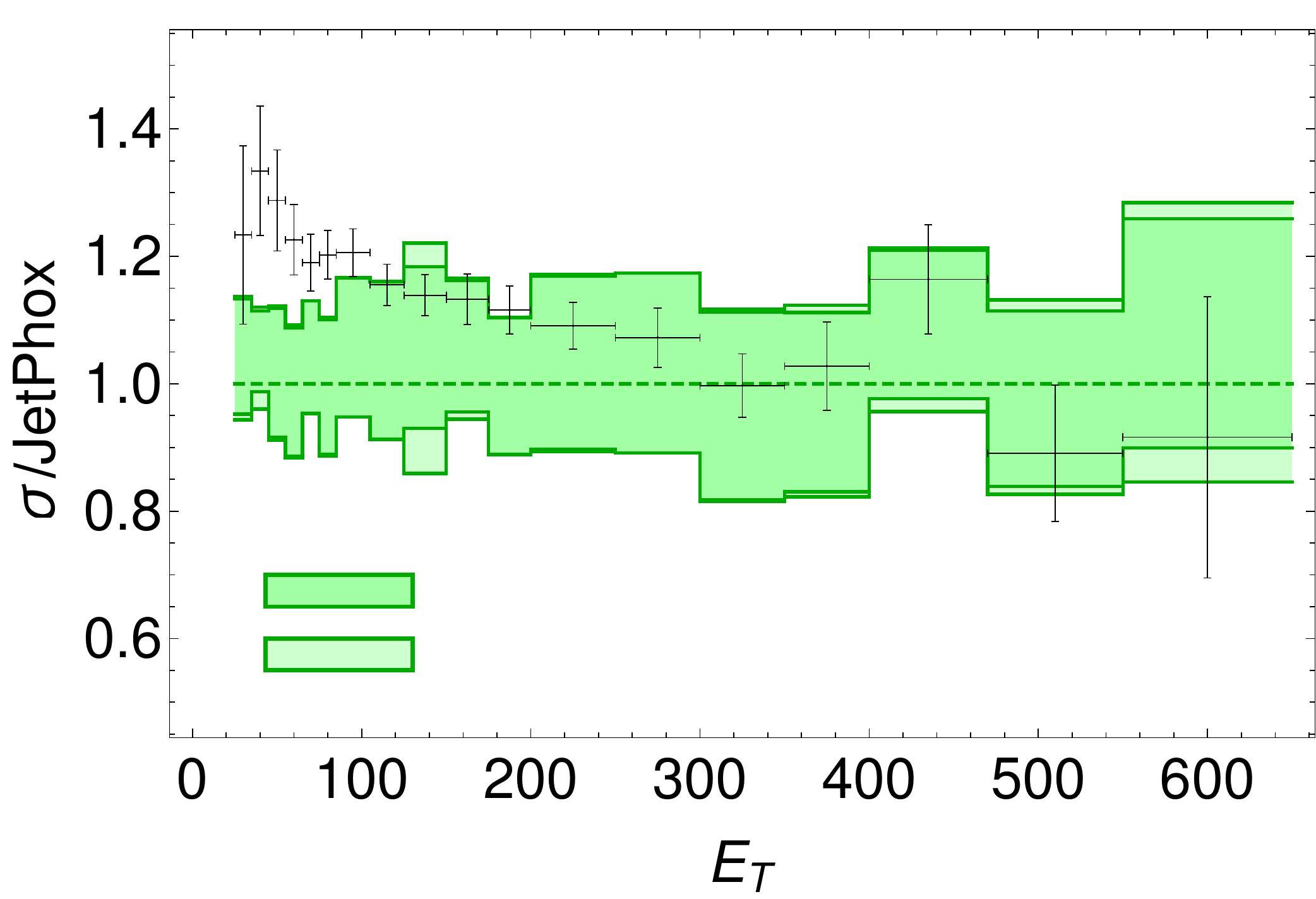}};
	\node[above,scale=1.0] at ($(c1)+(dc1)$) {$|\eta| < 0.6$};
	\node[right,scale=0.6] at ($(c1)+(dc2)$) {JetPhox (NLO unc.)};
	\node[right,scale=0.6] at ($(c1)+(dc3)$) {JetPhox (NLO + PDF unc.)};
%	\node[above,scale=1] at ($(c1)+(-1.5,-1.5)$) {$\alpha_e= \frac{1}{137}$};
%
	\node[above,scale=1.0] at ($(c2)+(dc1)$) {$0.6 < |\eta| < 1.37$};
	\node[right,scale=0.6] at ($(c2)+(dc2)$) {JetPhox (NLO unc.)};
	\node[right,scale=0.6] at ($(c2)+(dc3)$) {JetPhox (NLO + PDF unc.)};
%	\node[above,scale=1] at ($(c2)+(-1.5,-1.5)$) {$\alpha_e= \frac{1}{137}$};
%
	\node[above,scale=1.0] at ($(c3)+(dc1)$) {$1.56 < |\eta| < 1.81$};
	\node[right,scale=0.6] at ($(c3)+(dc2)$) {JetPhox (NLO unc.)};
	\node[right,scale=0.6] at ($(c3)+(dc3)$) {JetPhox (NLO + PDF unc.)};
%	\node[above,scale=1] at ($(c3)+(-1.5,-1.5)$) {$\alpha_e= \frac{1}{137}$};
%
	\node[above,scale=1.0] at ($(c4)+(dc1)$) {$1.81 < |\eta| < 2.37$};
	\node[right,scale=0.6] at ($(c4)+(dc2)$) {JetPhox (NLO unc.)};
	\node[right,scale=0.6] at ($(c4)+(dc3)$) {JetPhox (NLO + PDF unc.)};
%	\node[above,scale=1] at ($(c4)+(-1.5,-1.5)$) {$\alpha_e= \frac{1}{137}$};
\end{tikzpicture}
\caption{Comparison between the prediction from \jetphox and ATLAS data (black). Darker bands are scale uncertainties,
lighter bands also include PDF uncertainty. These plots use \jetphox default $\alpha_e = \frac{1}{137}$.}
\label{fig:compare0}
\end{center}
\end{figure}

  \begin{figure}[t]
\begin{center}
\begin{tikzpicture}
 	\coordinate (c1) at (-1,0);
 	\coordinate (c2) at (8,0);
 	\coordinate (c3) at (-1,-6);
 	\coordinate (c4) at (8,-6);
 	\coordinate (dc1) at (0.5,1.8);
        \coordinate (dc2) at (-1.5,-0.9);
        \coordinate (dc3) at (-1.5,-1.3);
        \coordinate (dc4) at (2.5,-1.5);
        \node at (c1) {\includegraphics[width=0.5\columnwidth]{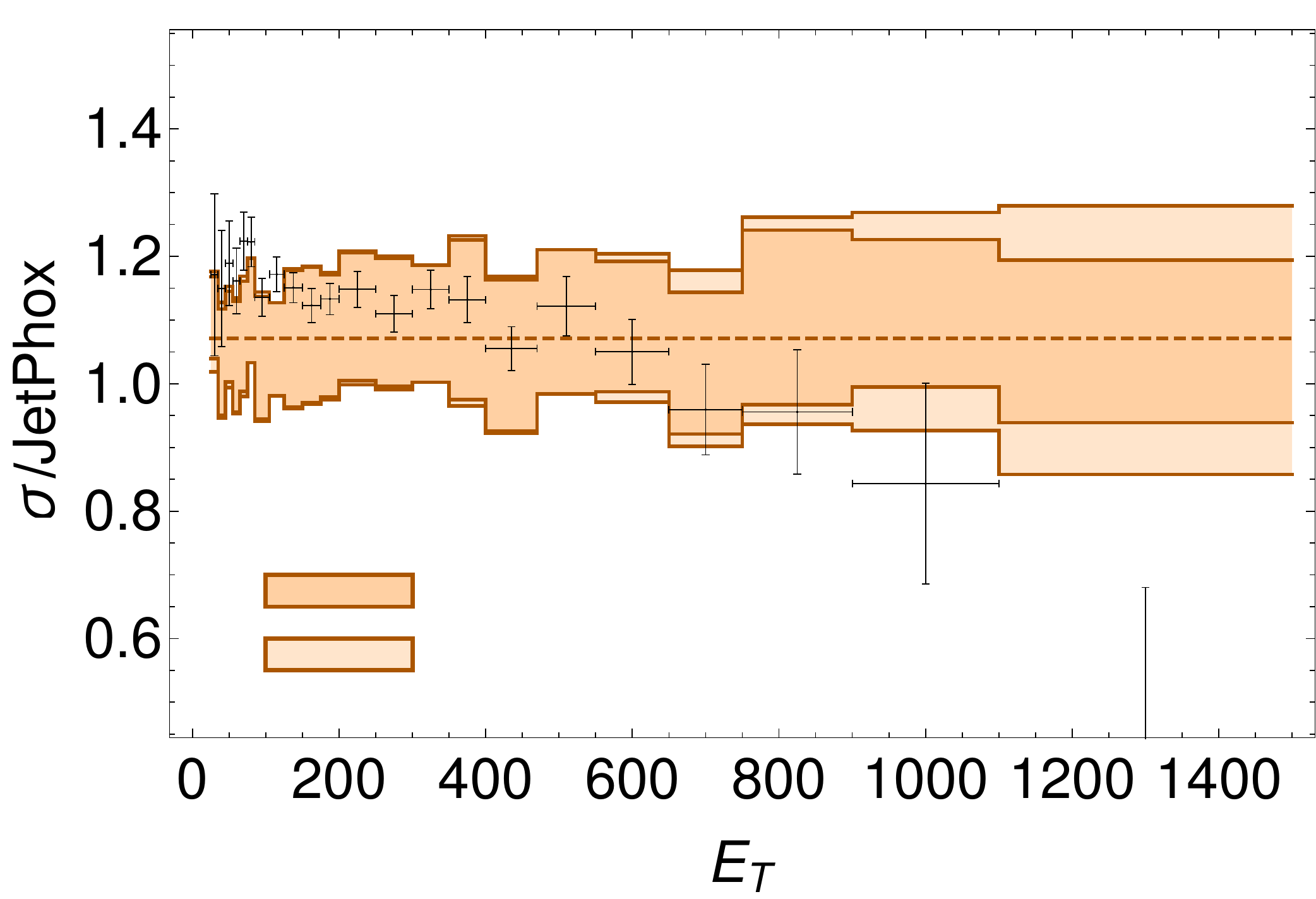}};
        \node at (c2) {\includegraphics[width=0.5\columnwidth]{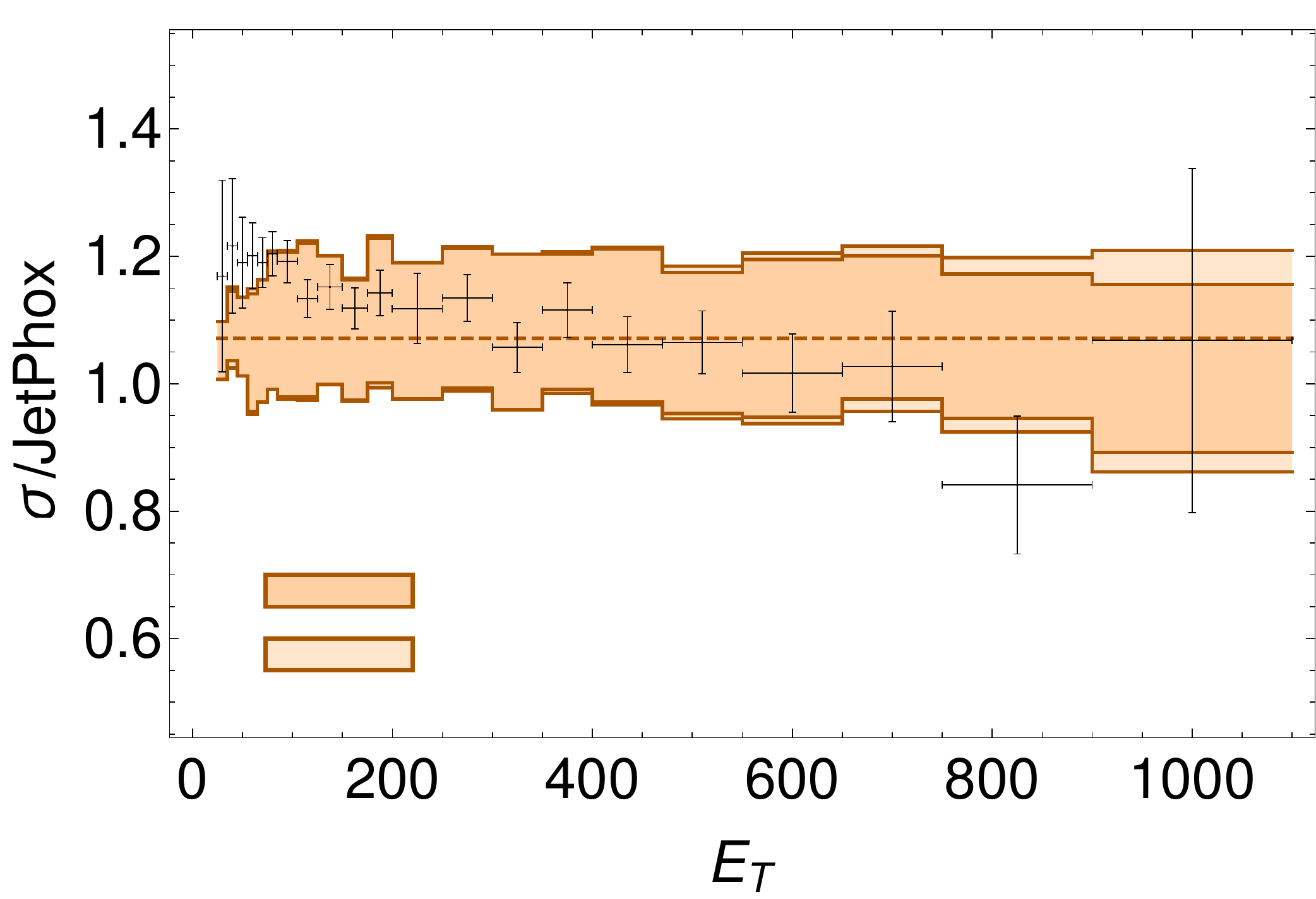}};
        \node at (c3) {\includegraphics[width=0.5\columnwidth]{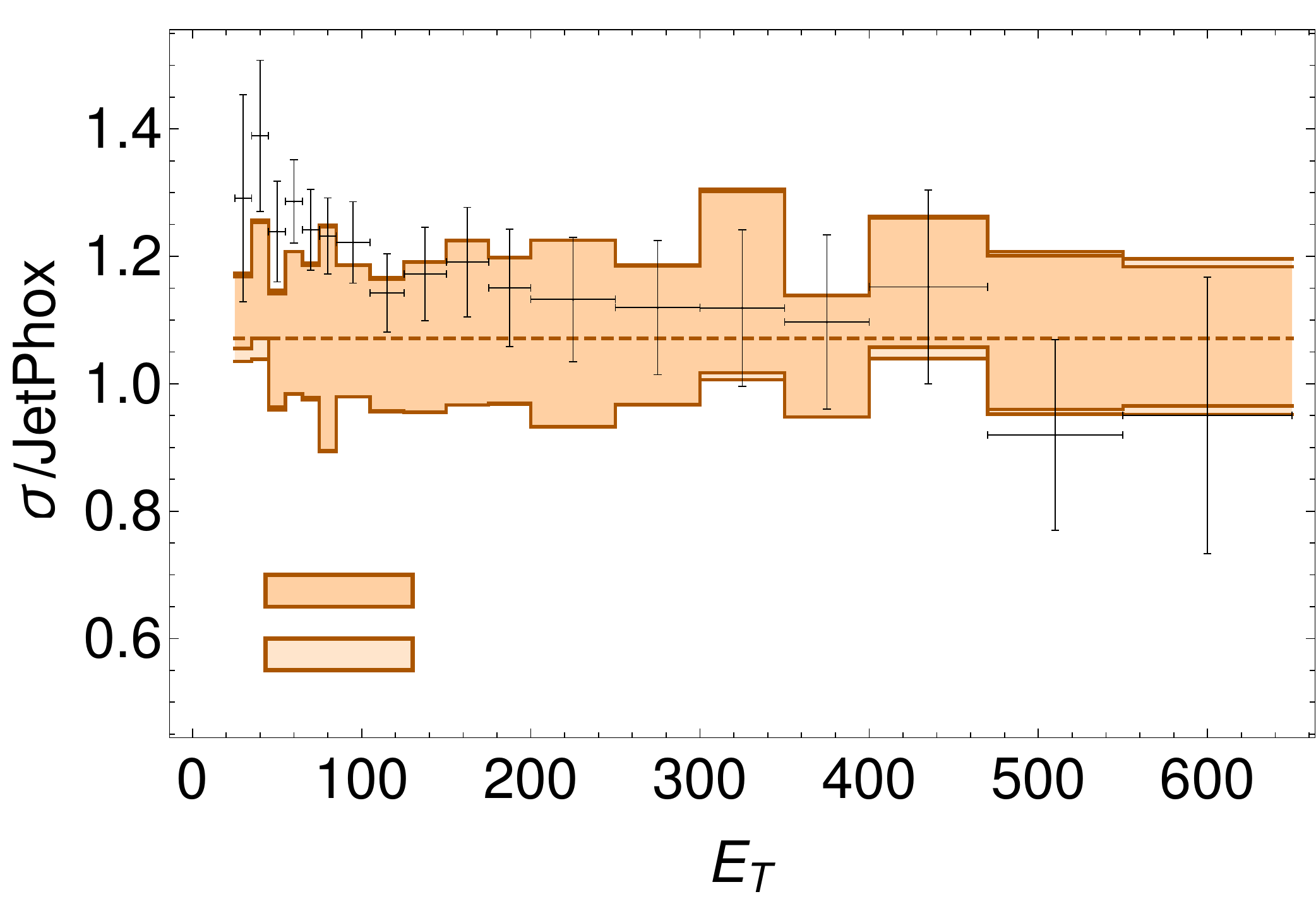}};
        \node at (c4) {\includegraphics[width=0.5\columnwidth]{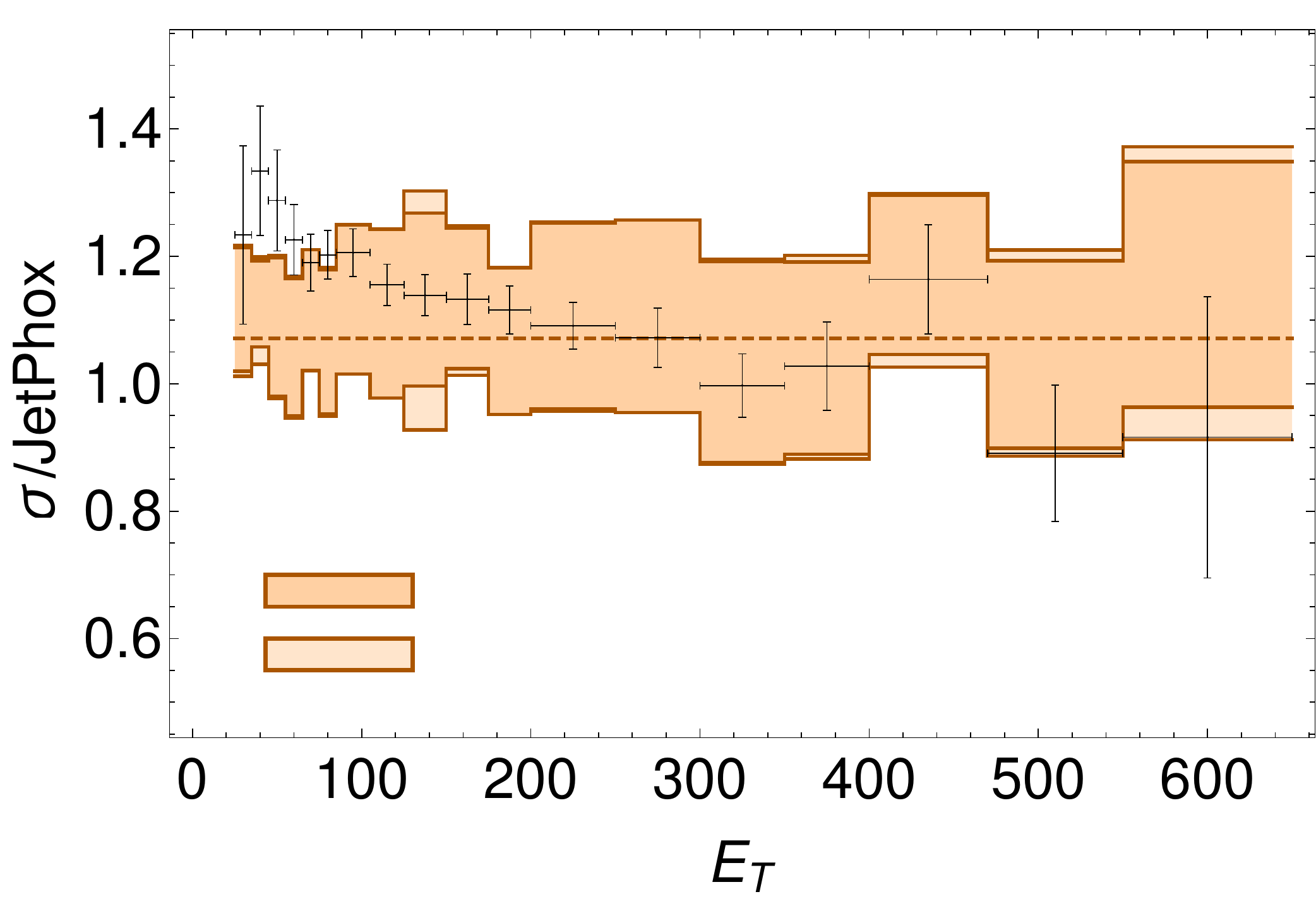}};
	\node[above,scale=1.0] at ($(c1)+(dc1)$) {$|\eta| < 0.6$};
	\node[right,scale=0.6] at ($(c1)+(dc2)$) {JetPhox + $\alpha_e$ (NLO unc.)};
	\node[right,scale=0.6] at ($(c1)+(dc3)$) {JetPhox + $\alpha_e$ (NLO + PDF unc.)};
%	\node[above,scale=1] at ($(c1)+(dc4)$) {\boxed{$\alpha_e= \frac{1}{129}$}};
%
	\node[above,scale=1.0] at ($(c2)+(dc1)$) {$0.6 < |\eta| < 1.37$};
	\node[right,scale=0.6] at ($(c2)+(dc2)$) {JetPhox + $\alpha_e$ (NLO unc.)};
	\node[right,scale=0.6] at ($(c2)+(dc3)$) {JetPhox + $\alpha_e$ (NLO + PDF unc.)};
	\node[above,scale=1.0] at ($(c3)+(dc1)$) {$1.56 < |\eta| < 1.81$};
	\node[right,scale=0.6] at ($(c3)+(dc2)$) {JetPhox + $\alpha_e$ (NLO unc.)};
	\node[right,scale=0.6] at ($(c3)+(dc3)$) {JetPhox + $\alpha_e$ (NLO + PDF unc.)};
	\node[above,scale=1.0] at ($(c4)+(dc1)$) {$1.81 < |\eta| < 2.37$};
	\node[right,scale=0.6] at ($(c4)+(dc2)$) {JetPhox + $\alpha_e$ (NLO unc.)};
	\node[right,scale=0.6] at ($(c4)+(dc3)$) {JetPhox + $\alpha_e$ (NLO + PDF unc.)};
\end{tikzpicture}
\caption{Comparison between the prediction from \jetphox with  $\alpha_e=\frac{1}{129}$ instead of JetPhox's default value of
$\alpha_e = \frac{1}{137}$ and ATLAS data.
Darker bands are scale uncertainties, lighter bands also include PDF uncertainty.}
\label{fig:compare1}
\end{center}
\end{figure}
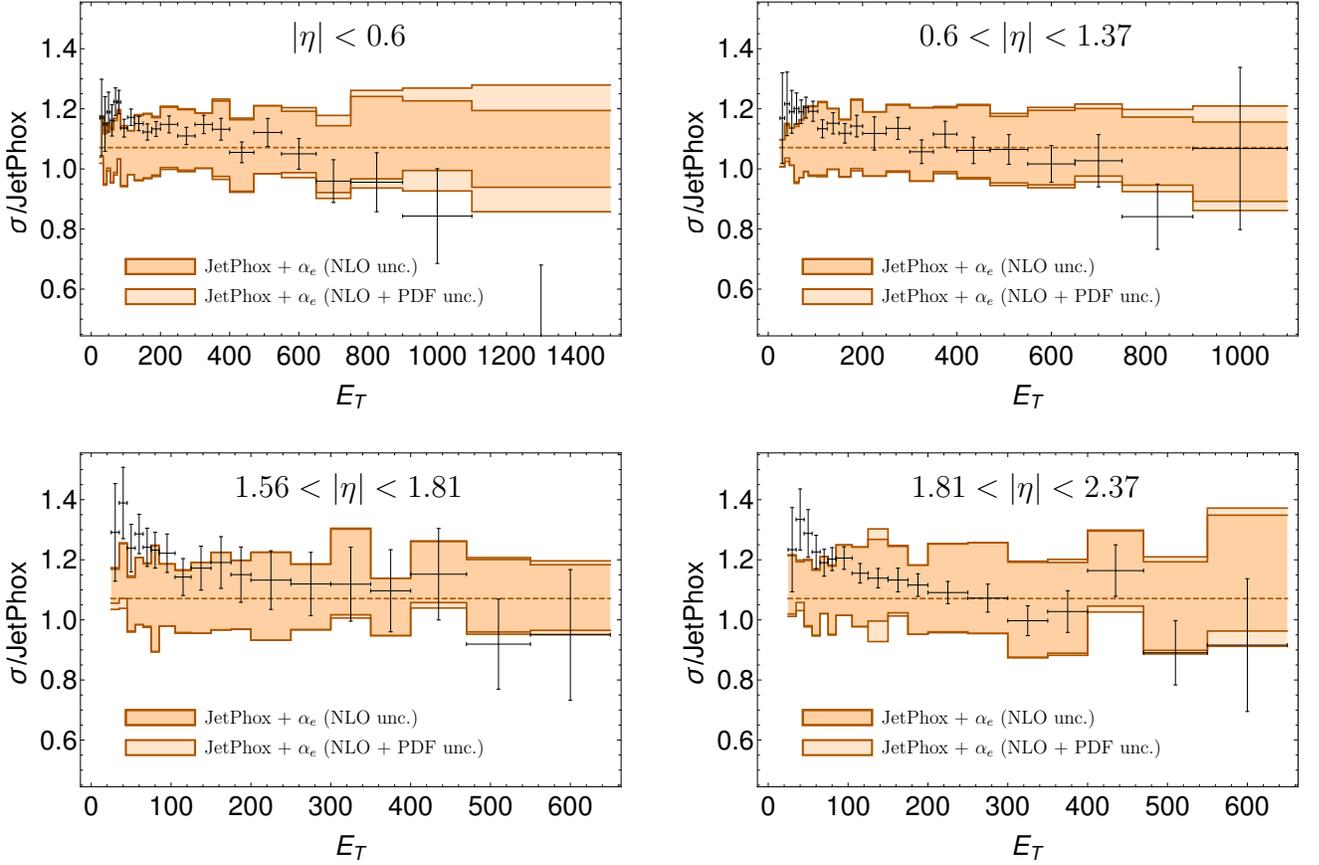

  \begin{figure}[t]
\begin{center}
\begin{tikzpicture}
 	\coordinate (c1) at (-1,0);
 	\coordinate (c2) at (8,0);
 	\coordinate (c3) at (-1,-6);
 	\coordinate (c4) at (8,-6);
 	\coordinate (dc1) at (0.5,1.8);
        \coordinate (dc2) at (-1.5,-0.9);
        \coordinate (dc3) at (-1.5,-1.3);
        \coordinate (dc4) at (2.5,-1.5);
        \node at (c1) {\includegraphics[width=0.5\columnwidth]{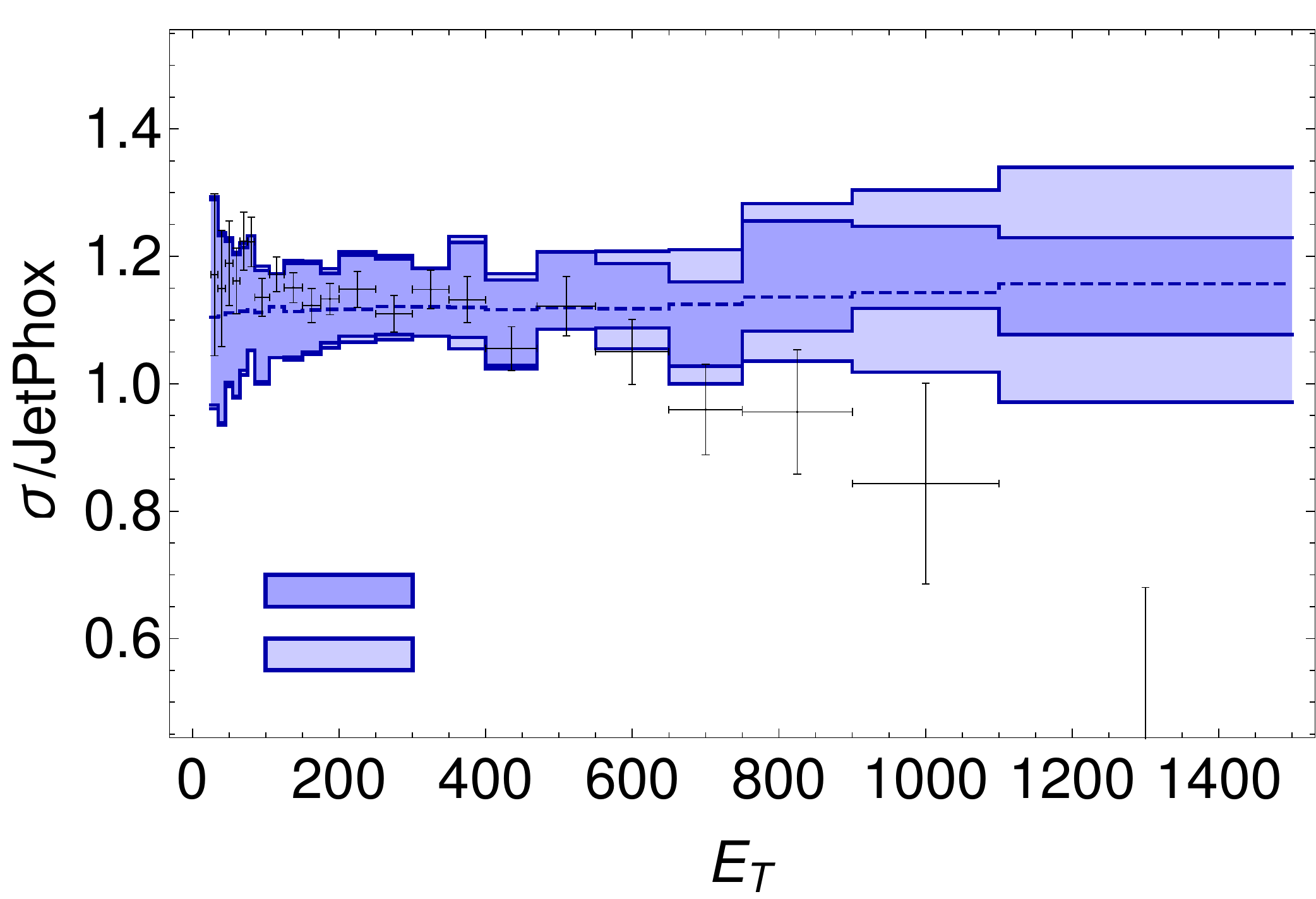}};
        \node at (c2) {\includegraphics[width=0.5\columnwidth]{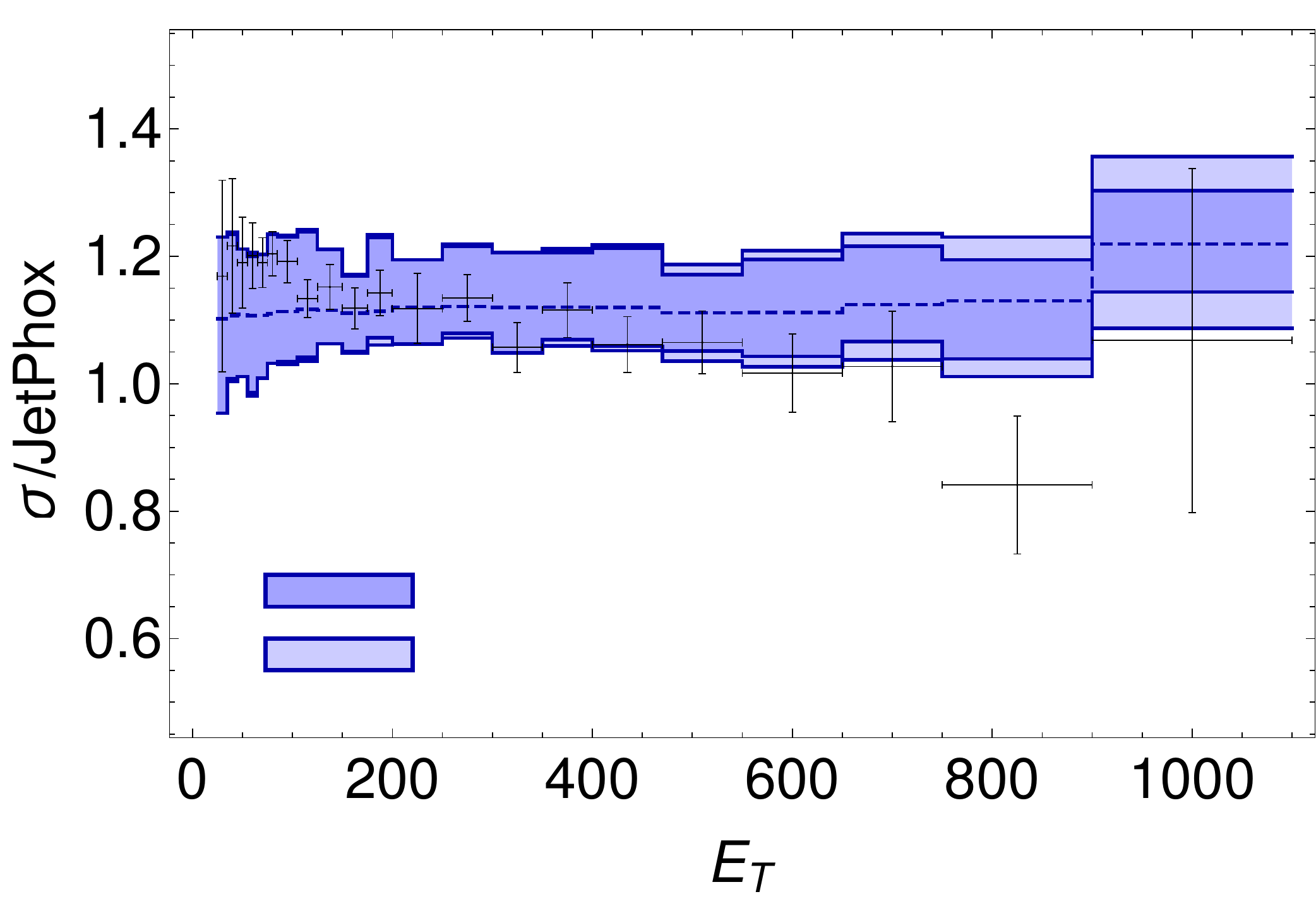}};
        \node at (c3) {\includegraphics[width=0.5\columnwidth]{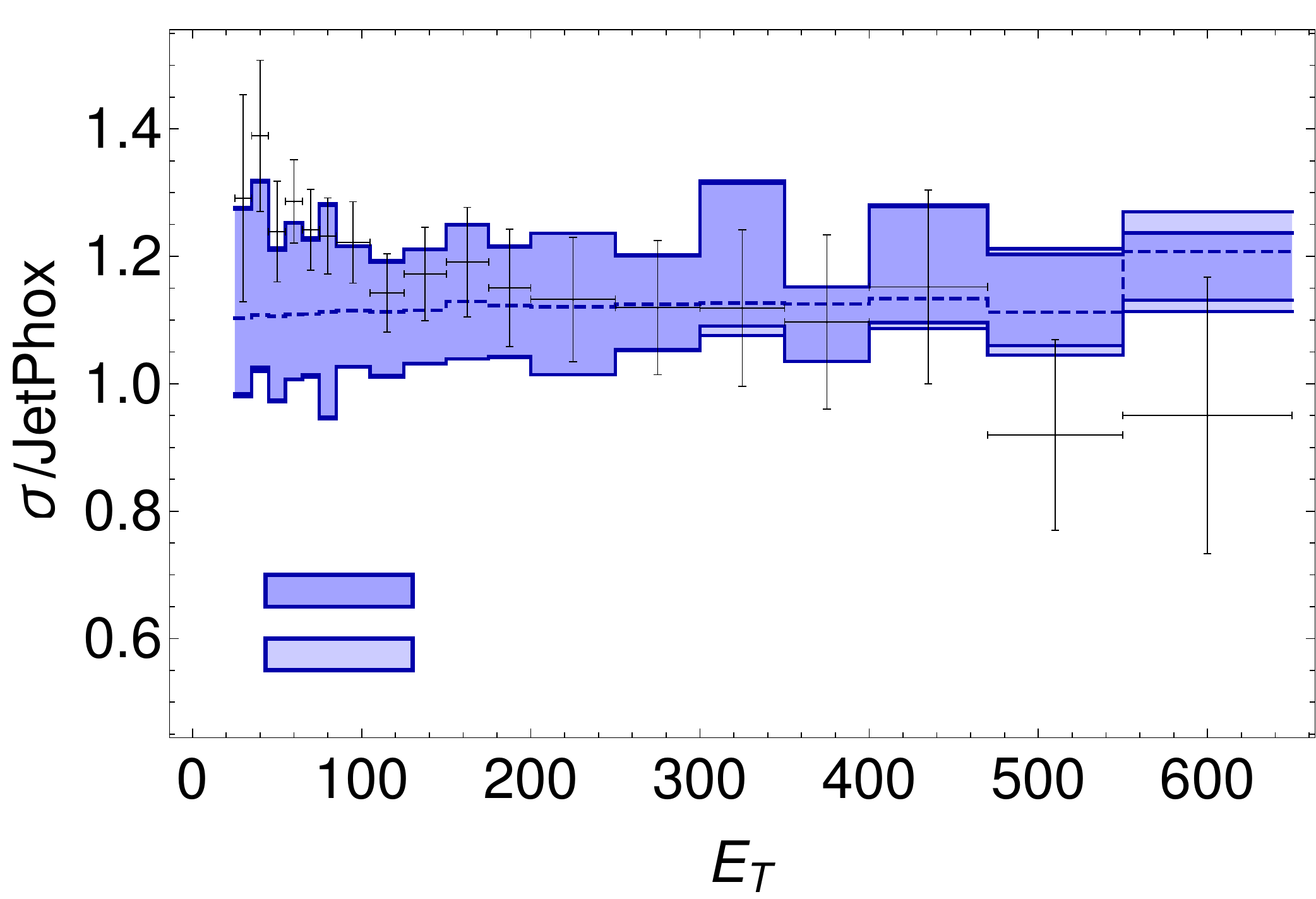}};
        \node at (c4) {\includegraphics[width=0.5\columnwidth]{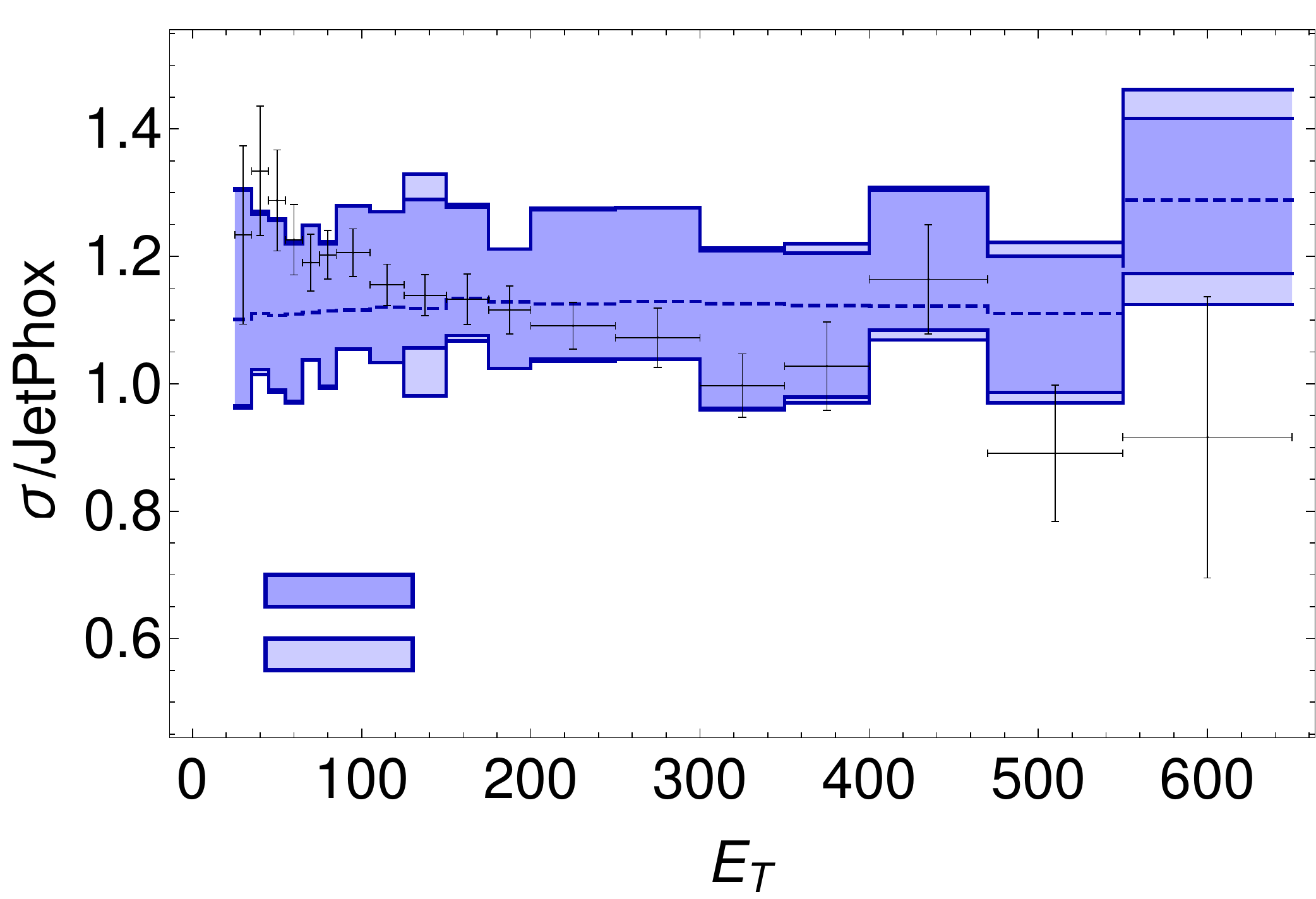}};
	\node[above,scale=1.0] at ($(c1)+(dc1)$) {$|\eta| < 0.6$};
	\node[right,scale=0.6] at ($(c1)+(dc2)$) {PeTeR (scale unc.)};
	\node[right,scale=0.6] at ($(c1)+(dc3)$) {PeTeR (scale + PDF unc.)};
	\node[above,scale=1.0] at ($(c2)+(dc1)$) {$0.6 < |\eta| < 1.37$};
	\node[right,scale=0.6] at ($(c2)+(dc2)$) {PeTeR (scale unc.)};
	\node[right,scale=0.6] at ($(c2)+(dc3)$) {PeTeR (scale + PDF unc.)};
	\node[above,scale=1.0] at ($(c3)+(dc1)$) {$1.56 < |\eta| < 1.81$};
	\node[right,scale=0.6] at ($(c3)+(dc2)$) {PeTeR (scale unc.)};
	\node[right,scale=0.6] at ($(c3)+(dc3)$) {PeTeR (scale + PDF unc.)};
	\node[above,scale=1.0] at ($(c4)+(dc1)$) {$1.81 < |\eta| < 2.37$};
	\node[right,scale=0.6] at ($(c4)+(dc2)$) {PeTeR (scale unc.)};
	\node[right,scale=0.6] at ($(c4)+(dc3)$) {PeTeR (scale + PDF unc.)};
\end{tikzpicture}
\caption{Comparison between the prediction from \peter matched to \jetphox and ATLAS data. 
Darker bands are scale uncertainties, lighter bands also include PDF uncertainty.}
\label{fig:compare2}
\end{center}
\end{figure}

  \begin{figure}[t]
\begin{center}
\begin{tikzpicture}
 	\coordinate (c1) at (-1,0);
 	\coordinate (c2) at (8,0);
 	\coordinate (c3) at (-1,-6);
 	\coordinate (c4) at (8,-6);
 	\coordinate (dc1) at (0.5,1.8);
        \coordinate (dc2) at (-1.5,-0.9);
        \coordinate (dc3) at (-1.5,-1.3);
        \coordinate (dc4) at (2.5,-1.5);
        \node at (c1) {\includegraphics[width=0.5\columnwidth]{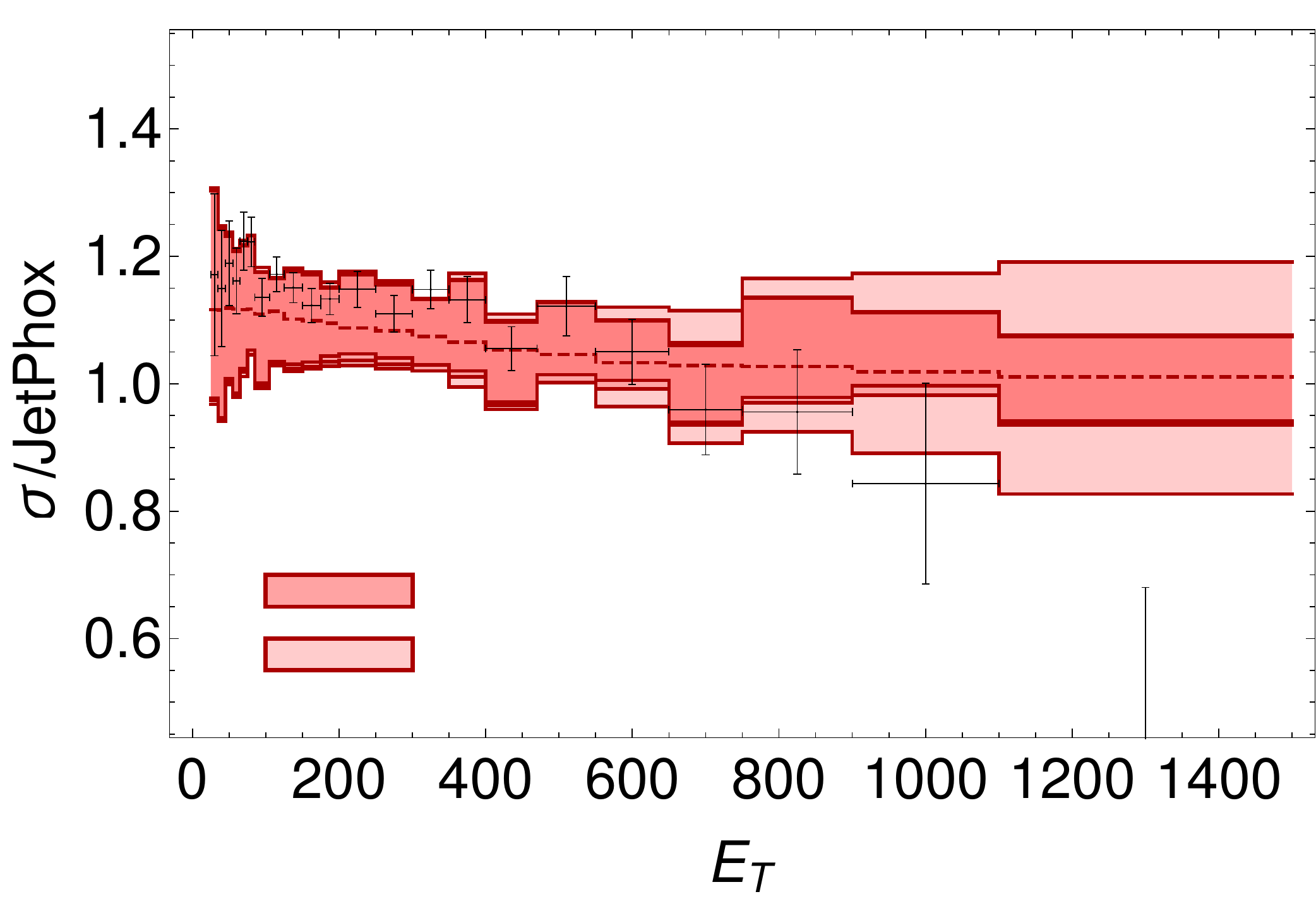}};
        \node at (c2) {\includegraphics[width=0.5\columnwidth]{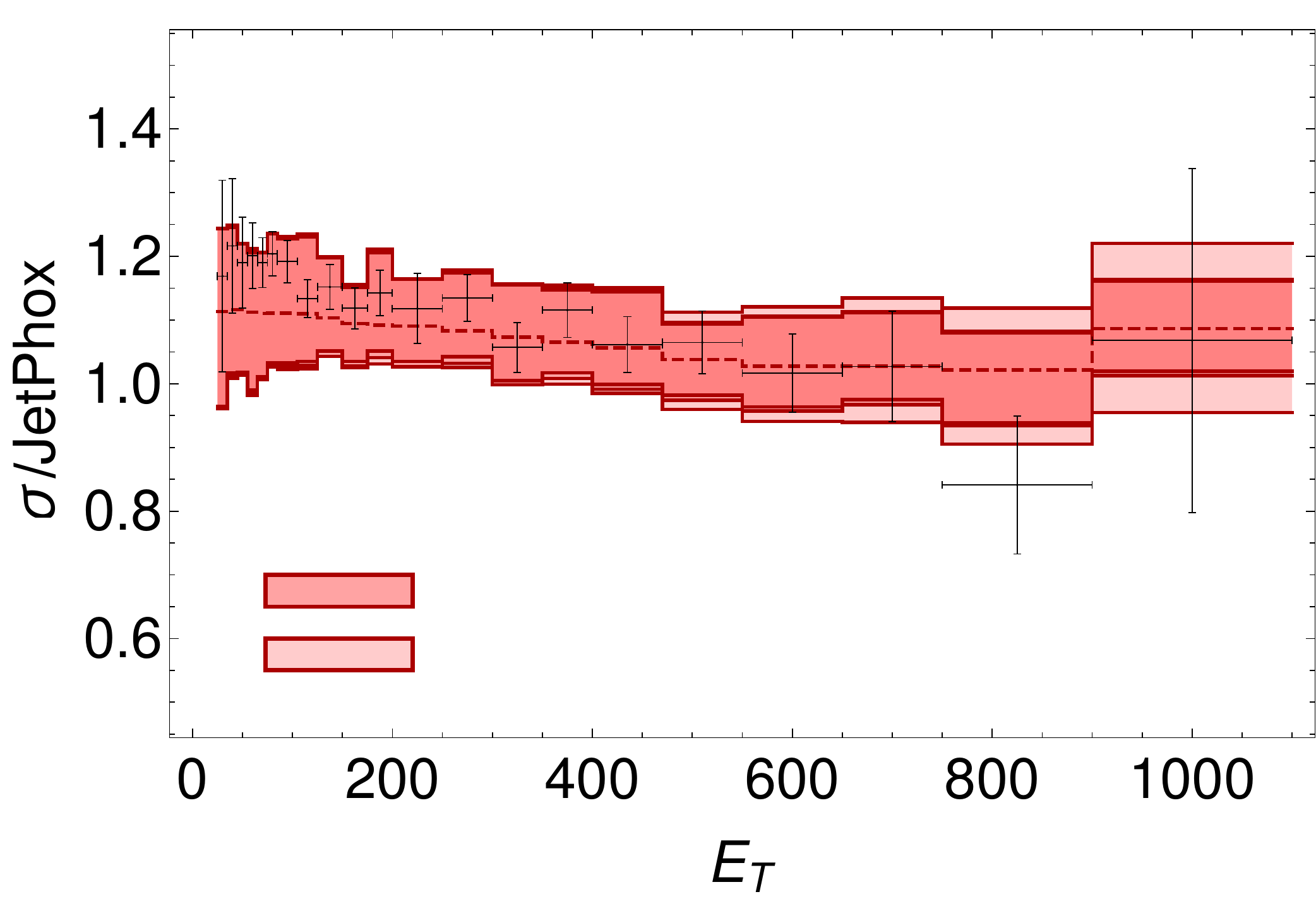}};
        \node at (c3) {\includegraphics[width=0.5\columnwidth]{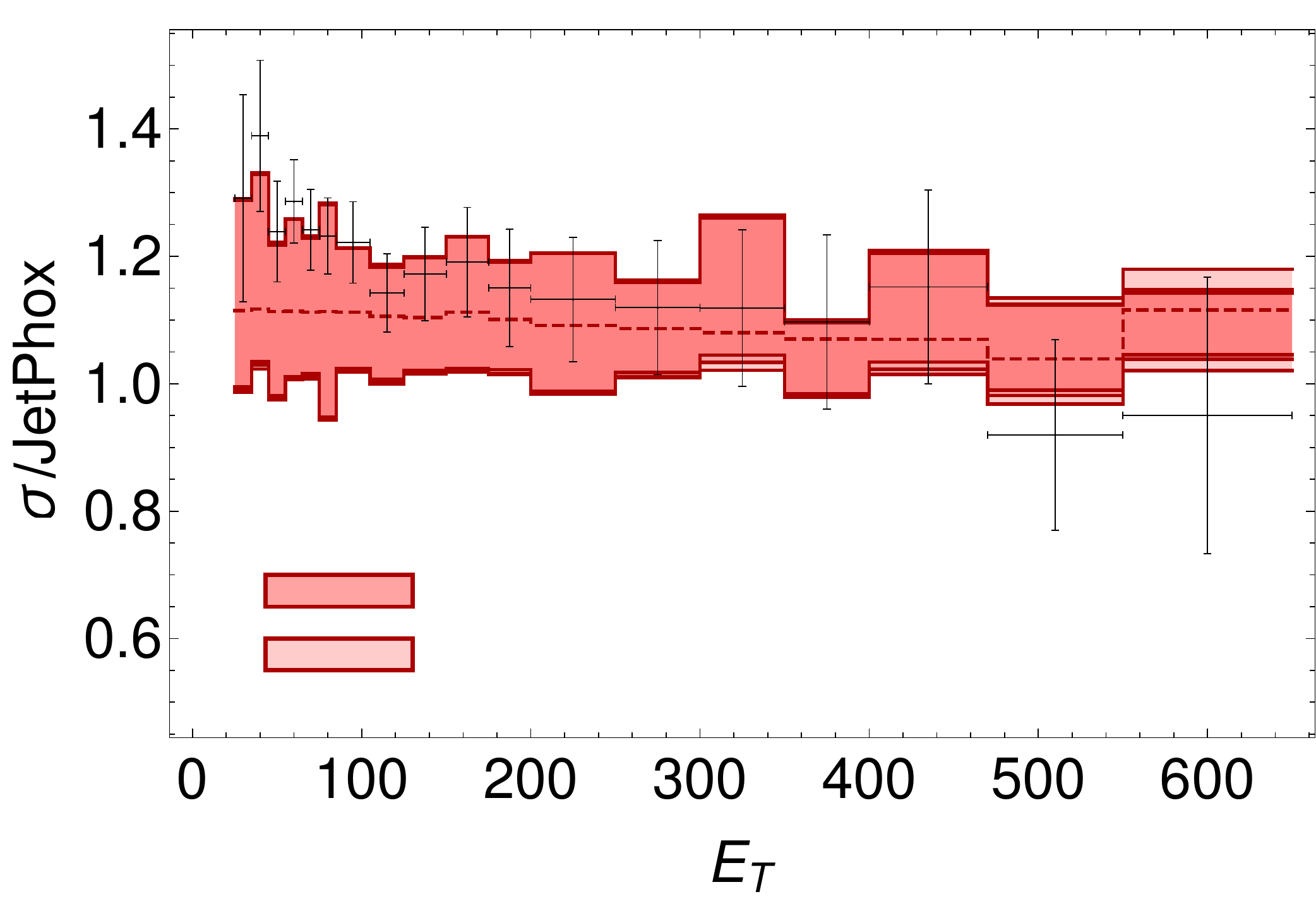}};
        \node at (c4) {\includegraphics[width=0.5\columnwidth]{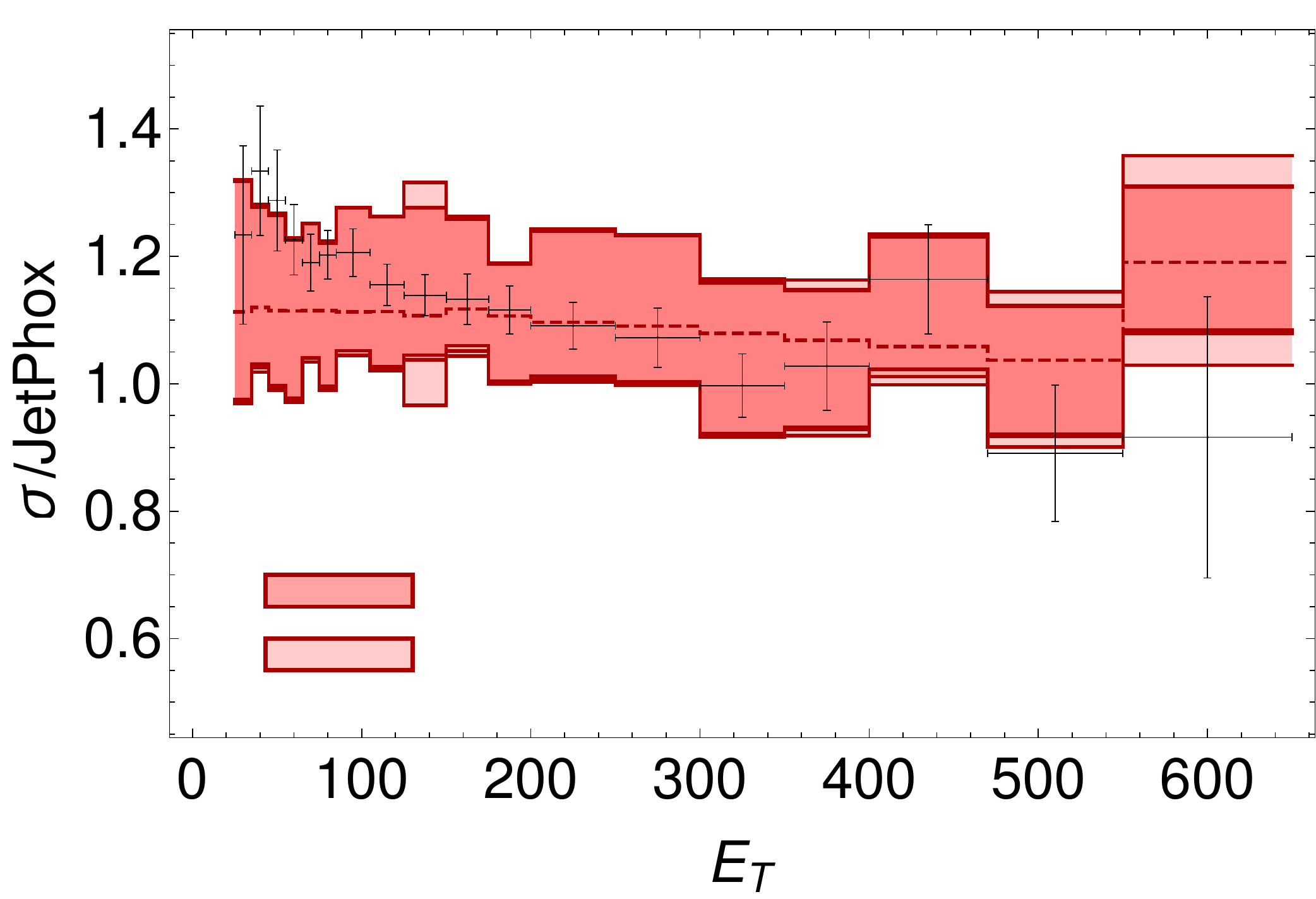}};
	\node[above,scale=1.0] at ($(c1)+(dc1)$) {$|\eta| < 0.6$};
	\node[right,scale=0.6] at ($(c1)+(dc2)$) {PeTeR + EW (scale unc.)};
	\node[right,scale=0.6] at ($(c1)+(dc3)$) {PeTeR + EW (scale + PDF unc.)};
	\node[above,scale=1.0] at ($(c2)+(dc1)$) {$0.6 < |\eta| < 1.37$};
	\node[right,scale=0.6] at ($(c2)+(dc2)$) {PeTeR + EW (scale unc.)};
	\node[right,scale=0.6] at ($(c2)+(dc3)$) {PeTeR + EW (scale + PDF unc.)};
	\node[above,scale=1.0] at ($(c3)+(dc1)$) {$1.56 < |\eta| < 1.81$};
	\node[right,scale=0.6] at ($(c3)+(dc2)$) {PeTeR + EW (scale unc.)};
	\node[right,scale=0.6] at ($(c3)+(dc3)$) {PeTeR + EW (scale + PDF unc.)};
	\node[above,scale=1.0] at ($(c4)+(dc1)$) {$1.81 < |\eta| < 2.37$};
	\node[right,scale=0.6] at ($(c4)+(dc2)$) {PeTeR + EW (scale unc.)};
	\node[right,scale=0.6] at ($(c4)+(dc3)$) {PeTeR + EW (scale + PDF unc.)};
\end{tikzpicture}
\caption{Comparison between the prediction from PeTeR including electroweak corrections and ATLAS data. 
Darker bands are scale uncertainties, lighter bands also include PDF uncertainty.}
\label{fig:compare3}
\end{center}
\end{figure}

\section{Results and discussion}
The results for the theoretical predictions of the direct photon spectrum as compared to the 8 TeV ATLAS data are shown in Figs~\ref{fig:compare0}-\ref{fig:compare3}.
The experimental study normalized their comparison to data~\cite{Aad:2016xcr}, but we prefer
to normalize to theory. 
 Since the theory prediction at NLO is not statistically limited, it should be much smoother, and thus one can distinguish statistical fluctuations in the data from theoretical uncertainty. Unfortunately, the \jetphox results we use
  as the theory reference, which were provided by the authors of~\cite{Aad:2016xcr}, are not completely smooth. Nevertheless, normalizing to the NLO theory does illuminate  a number of interesting features of the theoretical predictions, as we now discuss. 

The first set of plots in Fig.~\ref{fig:compare0} shows the comparison to \jetphoxd. The agreement is not great, particularly at low $E_T$ where fragmentation is important.

The second set of plots in Fig.~\ref{fig:compare1} is again a comparison to \jetphoxd, but now with the fine structure constant taken to be $\alpha_e = \frac{1}{129}$
instead of the \jetphox default value of $\alpha_e = \frac{1}{137}$. Since the whole cross section is proportional to $\alpha_e$, this shifts the theory prediction up by around 6\%. Comparing Figs.~\ref{fig:compare0} and \ref{fig:compare1}, one can see a definite improvement with the larger value of $\alpha_e$. A discussion of why a high-scale $\alpha_e$ is more appropriate can be found in~\cite{Czarnecki:1998tn,Becher:2013zua,Becher:2015yea}. 

The third set of plots, in Fig.~\ref{fig:compare2} shows the prediction from \peter matched to \jetphoxd. This theory prediction includes threshold resummation
at N${}^3$LL accuracy and is matched to the NLO fixed order results with fragmentation. The value of $\alpha_e$ used is the one from Fig.~\ref{fig:compare1}, $\alpha_e = \frac{1}{129}$. In going from Fig.~\ref{fig:compare1} to Fig.~\ref{fig:compare2} one can see an additional shift upward in the cross section. Looking at the central values of the prediction (dashed line), one sees that the increase is relatively larger at higher $E_T$. This is logical, as the resummation is more of an effect closer to threshold ($E_T \sim \ecm/2$) since the logarithms are larger.

Going from Fig.~\ref{fig:compare1} to Fig.~\ref{fig:compare2}  one also sees a tightening of the theory uncertainty band over most of the range  and a loosening of the band at small $E_T$. 
The change at high $E_T$ is due to the inclusion of higher order terms. The effect is most visible in the "factorization scale uncertainty" panel of Fig.~\ref{fig:unc}. At low $E_T$, the change is partly due to  \jetphox underestimating the theory uncertainty in this region (one can see this underestimation
clearly in Fig.~\ref{fig:compare1}). In fact, the theory uncertainty at low $E_T$ is quite hard to estimate since it is dominated
by uncertainty on non-perturbative physics associated with hadronization. In addition, \peter
slightly overestimates the perturbative uncertainty by combining all 4 variations (hard, jet, soft and factorization scale) in quadrature, while the variations are in fact highly correlated. If it would prove valuable, an improved error estimate might come from understanding the fragmentation uncertainty better and including correlations of the scale uncertainties.

Finally, we add in the electroweak corrections to get to Fig.~\ref{fig:compare3}. The electroweak corrections have the effect of lowering the cross section, particularly at high $E_T$. In all rapidity regions, this seems to produce improved agreement with data compared to \peter alone.  This provides one of the first direct demonstrations of the importance of the electroweak Sudakov logarithms in data. 

One application of the direct photon spectrum is to improve global PDF fits. In particular, at high $E_T$, the direct photon spectrum probes both  the quark and gluon PDFs at large $x$. This can be seen most clearly in the central region $\eta < |0.6|$ (top left panel in Fig.~\ref{fig:compare3}), where the PDF uncertainty is largest. The PDF uncertainty is smaller in the other rapidity regions because by kinetmatics alone, $x$ cannot be that large in the forward region. For example $\eta > 0.6$ implies that $x < 0.94$.

In conclusion, we have produced theory predictions for the direct photon $E_T$ spectrum incorporating fixed next-to-leading order results and fragmentation (through \jetphoxd), the resummation of threshold logarithms to next-to-next-to-next-to-leading logarithmic accuracy (through \peterd) and the leading electroweak Sudakov logarithms. Adding each successive theory contribution generates improved agreement with data. In particular, we see evidence for the importance of both higher-order QCD and electroweak contributions directly in the LHC data.

\section*{Acknowledgements}
This work began while the author was a Short Term Affiliate with the ATLAS collaboration. The author benefited from conversations with
S. Chekanov, T. Becher and X. Garcia i Tormo and would especially like to thank C. Lorentzen for writing \peterd. This research was supported in part by the U.S. Department of Energy under grant DE-SC0013607.

\newpage
\appendix
\section{Numerical predictions}

\begin{table}[ht!]
\centering
   {\small
 \renewcommand{\arraystretch}{1.3}
 \begin{tabular}{|c|ccccc|}
 \hline 
 $\ETg$ range [\GeV] &  $\sigma$ (\jetphox) & $\sigma$(\peter) & $\sigma$(\peter+EW) & PDF Unc. & [pb/\GeV] \\ \hline 
25--35	& $0.942\pm{}^{0.085}_{-0.028}$	& $0.971\pm{}^{0.16}_{-0.12}$	& $0.982\pm{}^{0.16}_{-0.13}$	& $\pm0.037$	& $\cdot 10^{3}$ \\ \hline
35--45	& $2.8\pm{}^{0.12}_{-0.32}$	& $2.9\pm{}^{0.33}_{-0.44}$	& $2.92\pm{}^{0.33}_{-0.45}$	& $\pm0.087$	& $\cdot 10^{2}$ \\ \hline
45--55	& $1.04\pm{}^{0.072}_{-0.066}$	& $1.07\pm{}^{0.11}_{-0.11}$	& $1.08\pm{}^{0.11}_{-0.11}$	& $\pm0.035$	& $\cdot 10^{2}$ \\ \hline
55--65	& $4.64\pm{}^{0.25}_{-0.5}$	& $4.82\pm{}^{0.39}_{-0.57}$	& $4.84\pm{}^{0.4}_{-0.59}$	& $\pm0.11$	& $\cdot 10^{1}$ \\ \hline
65--75	& $2.22\pm{}^{0.19}_{-0.17}$	& $2.31\pm{}^{0.21}_{-0.19}$	& $2.32\pm{}^{0.21}_{-0.2}$	& $\pm0.079$	& $\cdot 10^{1}$ \\ \hline
75--85	& $1.2\pm{}^{0.14}_{-0.043}$	& $1.25\pm{}^{0.13}_{-0.072}$	& $1.25\pm{}^{0.13}_{-0.08}$	& $\pm0.$	& $\cdot 10^{1}$ \\ \hline
85--105	& $6.01\pm{}^{0.37}_{-0.71}$	& $6.24\pm{}^{0.37}_{-0.61}$	& $6.22\pm{}^{0.37}_{-0.63}$	& $\pm0.17$	& $\cdot 10^{0}$ \\ \hline
105--125	& $2.32\pm{}^{0.12}_{-0.19}$	& $2.43\pm{}^{0.11}_{-0.17}$	& $2.41\pm{}^{0.12}_{-0.18}$	& $\pm0.$	& $\cdot 10^{0}$ \\ \hline
125--150	& $1.01\pm{}^{0.1}_{-0.1}$	& $1.05\pm{}^{0.072}_{-0.068}$	& $1.04\pm{}^{0.072}_{-0.073}$	& $\pm0.024$	& $\cdot 10^{0}$ \\ \hline
150--175	& $4.62\pm{}^{0.48}_{-0.43}$	& $4.81\pm{}^{0.32}_{-0.28}$	& $4.74\pm{}^{0.32}_{-0.31}$	& $\pm0.097$	& $\cdot 10^{-1}$ \\ \hline
175--200	& $2.21\pm{}^{0.21}_{-0.19}$	& $2.31\pm{}^{0.12}_{-0.11}$	& $2.26\pm{}^{0.12}_{-0.12}$	& $\pm0.061$	& $\cdot 10^{-1}$ \\ \hline
200--250	& $9.18\pm{}^{1.2}_{-0.57}$	& $9.57\pm{}^{0.73}_{-0.36}$	& $9.32\pm{}^{0.72}_{-0.44}$	& $\pm0.25$	& $\cdot 10^{-2}$ \\ \hline
250--300	& $3.3\pm{}^{0.39}_{-0.23}$	& $3.46\pm{}^{0.23}_{-0.14}$	& $3.34\pm{}^{0.23}_{-0.16}$	& $\pm0.085$	& $\cdot 10^{-2}$ \\ \hline
300--350	& $1.32\pm{}^{0.14}_{-0.085}$	& $1.38\pm{}^{0.075}_{-0.057}$	& $1.32\pm{}^{0.074}_{-0.066}$	& $\pm0.$	& $\cdot 10^{-2}$ \\ \hline
350--400	& $6.21\pm{}^{0.9}_{-0.56}$	& $6.49\pm{}^{0.59}_{-0.27}$	& $6.17\pm{}^{0.57}_{-0.31}$	& $\pm0.26$	& $\cdot 10^{-3}$ \\ \hline
400--470	& $2.88\pm{}^{0.25}_{-0.39}$	& $3.\pm{}^{0.13}_{-0.23}$	& $2.83\pm{}^{0.13}_{-0.24}$	& $\pm0.087$	& $\cdot 10^{-3}$ \\ \hline
470--550	& $1.08\pm{}^{0.14}_{-0.087}$	& $1.13\pm{}^{0.088}_{-0.034}$	& $1.05\pm{}^{0.084}_{-0.044}$	& $\pm0.$	& $\cdot 10^{-3}$ \\ \hline
550--650	& $4.13\pm{}^{0.47}_{-0.32}$	& $4.31\pm{}^{0.27}_{-0.12}$	& $3.98\pm{}^{0.26}_{-0.16}$	& $\pm0.21$	& $\cdot 10^{-4}$ \\ \hline
650--750	& $1.55\pm{}^{0.11}_{-0.22}$	& $1.63\pm{}^{0.051}_{-0.14}$	& $1.49\pm{}^{0.053}_{-0.14}$	& $\pm0.11$	& $\cdot 10^{-4}$ \\ \hline
750--900	& $4.89\pm{}^{0.78}_{-0.47}$	& $5.18\pm{}^{0.54}_{-0.25}$	& $4.69\pm{}^{0.5}_{-0.26}$	& $\pm0.39$	& $\cdot 10^{-5}$ \\ \hline
900--1100	& $11.9\pm{}^{1.7}_{-0.85}$	& $12.7\pm{}^{1.2}_{-0.27}$	& $11.3\pm{}^{1.}_{-0.41}$	& $\pm1.4$	& $\cdot 10^{-6}$ \\ \hline
1100--1500	& $15.6\pm{}^{1.8}_{-1.9}$	& $16.9\pm{}^{1.1}_{-1.2}$	& $14.8\pm{}^{0.96}_{-1.1}$	& $\pm2.4$	& $\cdot 10^{-7}$ \\ \hline
 \end{tabular}
}
 \caption{Predictions for bins in rapidity region $|\eta| < 0.6$. 
  \jetphox cross section and PDF uncertainties were provided by ATLAS.
  }
\label{tab:numbers_first}
\end{table}

\begin{table}[ht]
\centering
   {\small
 \renewcommand{\arraystretch}{1.3}
 \begin{tabular}{|c|ccccc|}
 \hline 
 $\ETg$ range [\GeV] &  $\sigma$ (\jetphox) & $\sigma$(\peter) & $\sigma$(\peter+EW) & PDF Unc. & [pb/\GeV] \\ \hline 
25--35	& $1.23\pm{}^{0.03}_{-0.074}$	& $1.26\pm{}^{0.15}_{-0.17}$	& $1.28\pm{}^{0.15}_{-0.17}$	& $\pm0.$	& $\cdot 10^{3}$ \\ \hline
35--45	& $3.42\pm{}^{0.24}_{-0.11}$	& $3.53\pm{}^{0.41}_{-0.31}$	& $3.56\pm{}^{0.41}_{-0.33}$	& $\pm0.095$	& $\cdot 10^{2}$ \\ \hline
45--55	& $1.3\pm{}^{0.078}_{-0.071}$	& $1.34\pm{}^{0.12}_{-0.12}$	& $1.35\pm{}^{0.13}_{-0.12}$	& $\pm0.$	& $\cdot 10^{2}$ \\ \hline
55--65	& $5.89\pm{}^{0.39}_{-0.63}$	& $6.09\pm{}^{0.51}_{-0.67}$	& $6.12\pm{}^{0.52}_{-0.69}$	& $\pm0.18$	& $\cdot 10^{1}$ \\ \hline
65--75	& $2.97\pm{}^{0.26}_{-0.28}$	& $3.07\pm{}^{0.26}_{-0.27}$	& $3.08\pm{}^{0.27}_{-0.29}$	& $\pm0.$	& $\cdot 10^{1}$ \\ \hline
75--85	& $1.57\pm{}^{0.2}_{-0.12}$	& $1.63\pm{}^{0.18}_{-0.11}$	& $1.63\pm{}^{0.18}_{-0.12}$	& $\pm0.$	& $\cdot 10^{1}$ \\ \hline
85--105	& $7.37\pm{}^{0.94}_{-0.63}$	& $7.66\pm{}^{0.8}_{-0.54}$	& $7.64\pm{}^{0.8}_{-0.58}$	& $\pm0.19$	& $\cdot 10^{0}$ \\ \hline
105--125	& $3.06\pm{}^{0.43}_{-0.26}$	& $3.19\pm{}^{0.35}_{-0.22}$	& $3.17\pm{}^{0.35}_{-0.23}$	& $\pm0.089$	& $\cdot 10^{0}$ \\ \hline
125--150	& $1.29\pm{}^{0.16}_{-0.087}$	& $1.35\pm{}^{0.12}_{-0.063}$	& $1.33\pm{}^{0.12}_{-0.073}$	& $\pm0.$	& $\cdot 10^{0}$ \\ \hline
150--175	& $5.9\pm{}^{0.51}_{-0.53}$	& $6.12\pm{}^{0.32}_{-0.33}$	& $6.03\pm{}^{0.32}_{-0.37}$	& $\pm0.1$	& $\cdot 10^{-1}$ \\ \hline
175--200	& $2.82\pm{}^{0.42}_{-0.18}$	& $2.93\pm{}^{0.3}_{-0.11}$	& $2.88\pm{}^{0.3}_{-0.13}$	& $\pm0.088$	& $\cdot 10^{-1}$ \\ \hline
200--250	& $1.2\pm{}^{0.13}_{-0.11}$	& $1.25\pm{}^{0.084}_{-0.064}$	& $1.22\pm{}^{0.083}_{-0.071}$	& $\pm0.$	& $\cdot 10^{-1}$ \\ \hline
250--300	& $4.07\pm{}^{0.54}_{-0.3}$	& $4.26\pm{}^{0.36}_{-0.16}$	& $4.11\pm{}^{0.35}_{-0.19}$	& $\pm0.099$	& $\cdot 10^{-2}$ \\ \hline
300--350	& $1.68\pm{}^{0.21}_{-0.18}$	& $1.76\pm{}^{0.13}_{-0.11}$	& $1.69\pm{}^{0.13}_{-0.12}$	& $\pm0.$	& $\cdot 10^{-2}$ \\ \hline
350--400	& $7.26\pm{}^{0.9}_{-0.54}$	& $7.59\pm{}^{0.58}_{-0.34}$	& $7.22\pm{}^{0.56}_{-0.39}$	& $\pm0.22$	& $\cdot 10^{-3}$ \\ \hline
400--470	& $3.11\pm{}^{0.41}_{-0.29}$	& $3.25\pm{}^{0.27}_{-0.18}$	& $3.06\pm{}^{0.26}_{-0.19}$	& $\pm0.087$	& $\cdot 10^{-3}$ \\ \hline
470--550	& $1.17\pm{}^{0.11}_{-0.13}$	& $1.21\pm{}^{0.066}_{-0.066}$	& $1.13\pm{}^{0.064}_{-0.069}$	& $\pm0.05$	& $\cdot 10^{-3}$ \\ \hline
550--650	& $4.02\pm{}^{0.47}_{-0.47}$	& $4.18\pm{}^{0.31}_{-0.26}$	& $3.86\pm{}^{0.3}_{-0.26}$	& $\pm0.19$	& $\cdot 10^{-4}$ \\ \hline
650--750	& $1.29\pm{}^{0.16}_{-0.11}$	& $1.36\pm{}^{0.11}_{-0.07}$	& $1.24\pm{}^{0.1}_{-0.073}$	& $\pm0.077$	& $\cdot 10^{-4}$ \\ \hline
750--900	& $3.77\pm{}^{0.36}_{-0.44}$	& $3.98\pm{}^{0.23}_{-0.32}$	& $3.59\pm{}^{0.21}_{-0.31}$	& $\pm0.27$	& $\cdot 10^{-5}$ \\ \hline
900--1100	& $7.25\pm{}^{0.58}_{-1.2}$	& $8.25\pm{}^{0.57}_{-0.51}$	& $7.36\pm{}^{0.52}_{-0.5}$	& $\pm0.74$	& $\cdot 10^{-6}$ \\ \hline
 \end{tabular}
   }
\caption{Predictions for bins in rapidity region $0.6 < |\eta| < 1.37$. }
\label{tab:numbers_first}
\end{table}

\begin{table}[ht]
\centering
   {\small
 \renewcommand{\arraystretch}{1.3}
 \begin{tabular}{|c|ccccc|}
 \hline 
 $\ETg$ range [\GeV] &  $\sigma$ (\jetphox) & $\sigma$(\peter) & $\sigma$(\peter+EW) & PDF Unc. & [pb/\GeV] \\ \hline 
25--35	& $3.67\pm{}^{0.33}_{-0.053}$	& $3.78\pm{}^{0.59}_{-0.41}$	& $3.82\pm{}^{0.59}_{-0.42}$	& $\pm0.11$	& $\cdot 10^{2}$ \\ \hline
35--45	& $1.03\pm{}^{0.18}_{0.}$	& $1.07\pm{}^{0.2}_{-0.079}$	& $1.08\pm{}^{0.2}_{-0.085}$	& $\pm0.032$	& $\cdot 10^{2}$ \\ \hline
45--55	& $4.17\pm{}^{0.27}_{-0.42}$	& $4.3\pm{}^{0.4}_{-0.51}$	& $4.33\pm{}^{0.41}_{-0.53}$	& $\pm0.11$	& $\cdot 10^{1}$ \\ \hline
55--65	& $1.79\pm{}^{0.23}_{-0.15}$	& $1.85\pm{}^{0.24}_{-0.17}$	& $1.86\pm{}^{0.24}_{-0.18}$	& $\pm0.$	& $\cdot 10^{1}$ \\ \hline
65--75	& $0.923\pm{}^{0.099}_{-0.08}$	& $0.956\pm{}^{0.1}_{-0.083}$	& $0.958\pm{}^{0.1}_{-0.087}$	& $\pm0.022$	& $\cdot 10^{1}$ \\ \hline
75--85	& $5.02\pm{}^{0.82}_{-0.82}$	& $5.21\pm{}^{0.79}_{-0.78}$	& $5.21\pm{}^{0.79}_{-0.79}$	& $\pm0.12$	& $\cdot 10^{0}$ \\ \hline
85--105	& $2.36\pm{}^{0.25}_{-0.2}$	& $2.45\pm{}^{0.22}_{-0.19}$	& $2.45\pm{}^{0.22}_{-0.21}$	& $\pm0.$	& $\cdot 10^{0}$ \\ \hline
105--125	& $0.956\pm{}^{0.083}_{-0.1}$	& $0.994\pm{}^{0.069}_{-0.089}$	& $0.987\pm{}^{0.07}_{-0.093}$	& $\pm0.019$	& $\cdot 10^{0}$ \\ \hline
125--150	& $4.\pm{}^{0.45}_{-0.43}$	& $4.17\pm{}^{0.36}_{-0.31}$	& $4.12\pm{}^{0.36}_{-0.33}$	& $\pm0.$	& $\cdot 10^{-1}$ \\ \hline
150--175	& $1.7\pm{}^{0.24}_{-0.17}$	& $1.79\pm{}^{0.19}_{-0.14}$	& $1.77\pm{}^{0.19}_{-0.15}$	& $\pm0.$	& $\cdot 10^{-1}$ \\ \hline
175--200	& $8.36\pm{}^{0.99}_{-0.79}$	& $8.76\pm{}^{0.72}_{-0.62}$	& $8.59\pm{}^{0.71}_{-0.66}$	& $\pm0.13$	& $\cdot 10^{-2}$ \\ \hline
200--250	& $3.29\pm{}^{0.47}_{-0.43}$	& $3.44\pm{}^{0.36}_{-0.33}$	& $3.35\pm{}^{0.35}_{-0.33}$	& $\pm0.$	& $\cdot 10^{-2}$ \\ \hline
250--300	& $1.04\pm{}^{0.11}_{-0.1}$	& $1.1\pm{}^{0.074}_{-0.07}$	& $1.06\pm{}^{0.072}_{-0.074}$	& $\pm0.015$	& $\cdot 10^{-2}$ \\ \hline
300--350	& $3.6\pm{}^{0.78}_{-0.18}$	& $3.79\pm{}^{0.63}_{-0.12}$	& $3.63\pm{}^{0.61}_{-0.16}$	& $\pm0.12$	& $\cdot 10^{-3}$ \\ \hline
350--400	& $1.48\pm{}^{0.094}_{-0.17}$	& $1.56\pm{}^{0.037}_{-0.12}$	& $1.48\pm{}^{0.041}_{-0.13}$	& $\pm0.$	& $\cdot 10^{-3}$ \\ \hline
400--470	& $4.75\pm{}^{0.84}_{-0.06}$	& $5.03\pm{}^{0.64}_{-0.17}$	& $4.74\pm{}^{0.61}_{-0.21}$	& $\pm0.13$	& $\cdot 10^{-4}$ \\ \hline
470--550	& $1.48\pm{}^{0.18}_{-0.15}$	& $1.54\pm{}^{0.13}_{-0.073}$	& $1.44\pm{}^{0.12}_{-0.079}$	& $\pm0.057$	& $\cdot 10^{-4}$ \\ \hline
550--650	& $3.05\pm{}^{0.32}_{-0.3}$	& $3.44\pm{}^{0.083}_{-0.22}$	& $3.18\pm{}^{0.092}_{-0.22}$	& $\pm0.16$	& $\cdot 10^{-5}$ \\ \hline
 \end{tabular}
   }
\caption{Predictions for bins in rapidity region $1.56 < |\eta| < 1.81$. }
\label{tab:numbers_first}
\end{table}

\begin{table}[ht]
\centering
   {\small
 \renewcommand{\arraystretch}{1.3}
 \begin{tabular}{|c|ccccc|}
 \hline 
 $\ETg$ range [\GeV] &  $\sigma$ (\jetphox) & $\sigma$(\peter) & $\sigma$(\peter+EW) & PDF Unc. & [pb/\GeV] \\ \hline 
25--35	& $8.22\pm{}^{1.1}_{-0.39}$	& $8.45\pm{}^{1.6}_{-1.}$	& $8.54\pm{}^{1.6}_{-1.1}$	& $\pm0.23$	& $\cdot 10^{2}$ \\ \hline
35--45	& $2.28\pm{}^{0.26}_{-0.028}$	& $2.36\pm{}^{0.33}_{-0.19}$	& $2.38\pm{}^{0.33}_{-0.2}$	& $\pm0.081$	& $\cdot 10^{2}$ \\ \hline
45--55	& $0.865\pm{}^{0.1}_{-0.073}$	& $0.894\pm{}^{0.12}_{-0.095}$	& $0.9\pm{}^{0.12}_{-0.099}$	& $\pm0.022$	& $\cdot 10^{2}$ \\ \hline
55--65	& $3.91\pm{}^{0.34}_{-0.45}$	& $4.05\pm{}^{0.4}_{-0.5}$	& $4.07\pm{}^{0.41}_{-0.52}$	& $\pm0.095$	& $\cdot 10^{1}$ \\ \hline
65--75	& $1.94\pm{}^{0.25}_{-0.091}$	& $2.02\pm{}^{0.25}_{-0.13}$	& $2.02\pm{}^{0.25}_{-0.15}$	& $\pm0.$	& $\cdot 10^{1}$ \\ \hline
75--85	& $1.05\pm{}^{0.11}_{-0.12}$	& $1.09\pm{}^{0.1}_{-0.12}$	& $1.09\pm{}^{0.1}_{-0.12}$	& $\pm0.025$	& $\cdot 10^{1}$ \\ \hline
85--105	& $4.77\pm{}^{0.79}_{-0.25}$	& $4.97\pm{}^{0.73}_{-0.27}$	& $4.96\pm{}^{0.73}_{-0.31}$	& $\pm0.$	& $\cdot 10^{0}$ \\ \hline
105--125	& $1.9\pm{}^{0.3}_{-0.17}$	& $1.99\pm{}^{0.27}_{-0.15}$	& $1.98\pm{}^{0.27}_{-0.16}$	& $\pm0.$	& $\cdot 10^{0}$ \\ \hline
125--150	& $7.8\pm{}^{1.4}_{-0.54}$	& $8.14\pm{}^{1.2}_{-0.45}$	& $8.06\pm{}^{1.2}_{-0.5}$	& $\pm0.89$	& $\cdot 10^{-1}$ \\ \hline
150--175	& $3.14\pm{}^{0.51}_{-0.14}$	& $3.32\pm{}^{0.42}_{-0.17}$	& $3.27\pm{}^{0.41}_{-0.19}$	& $\pm0.097$	& $\cdot 10^{-1}$ \\ \hline
175--200	& $1.46\pm{}^{0.15}_{-0.16}$	& $1.54\pm{}^{0.11}_{-0.14}$	& $1.51\pm{}^{0.11}_{-0.15}$	& $\pm0.$	& $\cdot 10^{-1}$ \\ \hline
200--250	& $5.31\pm{}^{0.9}_{-0.55}$	& $5.58\pm{}^{0.73}_{-0.43}$	& $5.44\pm{}^{0.72}_{-0.44}$	& $\pm0.13$	& $\cdot 10^{-2}$ \\ \hline
250--300	& $1.42\pm{}^{0.25}_{-0.15}$	& $1.49\pm{}^{0.19}_{-0.12}$	& $1.44\pm{}^{0.19}_{-0.12}$	& $\pm0.$	& $\cdot 10^{-2}$ \\ \hline
300--350	& $4.49\pm{}^{0.51}_{-0.82}$	& $4.72\pm{}^{0.35}_{-0.69}$	& $4.52\pm{}^{0.34}_{-0.67}$	& $\pm0.12$	& $\cdot 10^{-3}$ \\ \hline
350--400	& $1.41\pm{}^{0.16}_{-0.24}$	& $1.48\pm{}^{0.11}_{-0.19}$	& $1.4\pm{}^{0.1}_{-0.18}$	& $\pm0.067$	& $\cdot 10^{-3}$ \\ \hline
400--470	& $3.56\pm{}^{0.75}_{-0.083}$	& $3.73\pm{}^{0.61}_{-0.13}$	& $3.52\pm{}^{0.57}_{-0.16}$	& $\pm0.12$	& $\cdot 10^{-4}$ \\ \hline
470--550	& $8.62\pm{}^{0.98}_{-1.4}$	& $8.94\pm{}^{0.72}_{-1.}$	& $8.35\pm{}^{0.69}_{-0.96}$	& $\pm0.53$	& $\cdot 10^{-5}$ \\ \hline
550--650	& $1.26\pm{}^{0.33}_{-0.13}$	& $1.52\pm{}^{0.15}_{-0.14}$	& $1.4\pm{}^{0.14}_{-0.13}$	& $\pm0.14$	& $\cdot 10^{-5}$ \\ \hline
 \end{tabular}
   }
\caption{Predictions for bins in rapidity region $1.81 < |\eta| < 2.37$. }
\label{tab:numbers_four}
\end{table}
\FloatBarrier

\bibliography{photon}

\bibliographystyle{utphys}

\end{document}